\documentclass[a4paper,12pt]{article}
\usepackage{amssymb}
\usepackage{amsmath}
\usepackage{fullpage}
\usepackage{boxedminipage}
\usepackage{listings}
\usepackage{minitoc}
\makeatletter \@addtoreset{equation}{section} \makeatother

\begin{document}

\begin{titlepage}

    \thispagestyle{empty}
    \begin{flushright}
        \hfill{CERN-PH-TH/2007-025} \\
    \end{flushright}

    \vspace{5pt}
    \begin{center}
        { \Huge{\textbf{Attractor Horizon Geometries\\\vspace{15pt}of Extremal Black Holes}}}\vspace{25pt}
        \vspace{55pt}

        {\bf Stefano Bellucci$^\clubsuit$, Sergio Ferrara$^{\diamondsuit\clubsuit\flat}$ and\ Alessio Marrani$^{\heartsuit\clubsuit}$}

        \vspace{15pt}

        {$\clubsuit$ \it INFN - Laboratori Nazionali di Frascati, \\
        Via Enrico Fermi 40,00044 Frascati, Italy\\
        \texttt{bellucci,marrani@lnf.infn.it}}

        \vspace{10pt}

        {$\diamondsuit$ \it Physics Department,Theory Unit, CERN, \\
        CH 1211, Geneva 23, Switzerland\\
        \texttt{sergio.ferrara@cern.ch}}

        \vspace{10pt}
         {$\flat$ \it Department of Physics and Astronomy,\\
        University of California, Los Angeles, CA USA\\
        \texttt{ferrara@physics.ucla.edu}}

         \vspace{10pt}

        {$\heartsuit$ \it Museo Storico della Fisica e\\
        Centro Studi e Ricerche ``Enrico Fermi"\\
        Via Panisperna 89A, 00184 Roma, Italy}

        \vspace{50pt}
        \noindent \textit{Contribution to the Proceedings of the XVII SIGRAV Conference,\\4--7 September 2006, Turin, Italy}
\end{center}


\begin{abstract}
We report on recent advances in the study of critical points of the
``black hole effective potential'' $V_{BH}$ (usually named
\textit{attractors}) of $\mathcal{N}=2$, $d=4$ supergravity coupled
to $n_{V} $ Abelian vector multiplets, in an asymptotically flat
extremal black hole background described by $2n_{V}+2$ dyonic
charges and (complex) scalar fields which are coordinates of an
$n_{V}$-dimensional Special K\"{a}hler manifold.
\end{abstract}

\end{titlepage}
\newpage\baselineskip6 mm
\tableofcontents
\section{Introduction\label{Intro}}

After some seminal papers
\cite{FKS}-\nocite{Strom,FK1,FK2}\cite{FGK} of some years ago,
extremal black hole (BH) attractors have been recently widely
investigated \cite{Sen-old1}-\nocite
{GIJT,Sen-old2,K1,TT,G,GJMT,Ebra1,K2,Ira1,Tom,BFM,AoB-book,Lust1,Sen1,FKlast,Ebra2,FG2,BFGM1,rotating-attr,K3,Misra1,Lust2,Morales,BFMY,CYY,MRS,CdWMa}
\cite{DFT07-1}. Such a \textit{renaissance} is mainly due to the
(re)discovery of new classes of solutions to the attractor equations
corresponding to non-BPS (Bogomol'ny-Prasad-Sommerfeld) horizon
geometries.

An horizon extremal BH attractor geometry is in general supported by
particular configurations of the $1\times \left( 2n_{V}+2\right) $
symplectic vector of the BH field-strength fluxes, \textit{i.e.} of the BH
magnetic and electric charges:
\begin{equation}
\widetilde{\Gamma }\equiv \left( p^{\Lambda },q_{\Lambda }\right)
,~~~~~p^{\Lambda }\equiv \frac{1}{4\pi }\int_{S_{\infty }^{2}}\mathcal{F}%
^{\Lambda },~~~q_{\Lambda }\equiv \frac{1}{4\pi }\int_{S_{\infty }^{2}}%
\mathcal{G}_{\Lambda },~~~\Lambda =0,1,...,n_{V},  \label{Gamma-tilde}
\end{equation}
where, in the case of $\mathcal{N}=2$, $d=4$ Maxwell-Einstein supergravity
theories (MESGTs), $n_{V}$ denotes the number of Abelian vector
supermultiplets coupled to the supergravity one (containing the Maxwell
vector $A^{0}$, usually named \textit{graviphoton}). Here $\mathcal{F}%
^{\Lambda }=dA^{\Lambda }$ and $\mathcal{G}_{\Lambda }$ is the ``dual''
field-strength two-form \cite{3,4}.

In the present brief review we will consider only \textit{non-degenerate }($%
\frac{1}{2}$-BPS as well as non-BPS) geometries, \textit{i.e.} geometries
yielding a finite, non-vanishing horizon area, corresponding to the
so-called \textit{``large''} BHs. Due to the well-known Attractor Mechanism
\cite{FKS}-\nocite{Strom,FK1,FK2}\cite{FGK}, such BH horizon geometries are
actually \textit{critical}, because their Bekenstein-Hawking entropy \cite
{BH1}\textbf{\ }can be obtained by extremizing a properly defined,
positive-definite ``effective BH potential'' function $V_{BH}\left( \phi ,%
\widetilde{\Gamma }\right) $, where ``$\phi $'' denotes the set of real
scalars relevant for the Attractor Mechanism.

In $\mathcal{N}=2$, $d=4$ MESGTs, non-degenerate attractor horizon
geometries correspond to BH solitonic states belonging to $\frac{1}{2}$-BPS
``short massive multiplets'' or to ``long massive multiplets'' violating the
BPS bound \cite{BPS}, respectively with\footnote{%
Here and in what follows, the subscript ``$H$'' will denote values at the BH
event horizon.}
\begin{equation}
\begin{array}{l}
\frac{1}{2}\text{\textit{-BPS}:~~}0<\left| Z\right| _{H}=M_{ADM,H}; \\
~ \\
\text{\textit{non-BPS~}}\left\{
\begin{array}{l}
Z\neq 0\text{:~~}0<\left| Z\right| _{H}<M_{ADM,H}; \\
\\
Z=0\text{:~~}0=\left| Z\right| _{H}<M_{ADM,H},
\end{array}
\right. \text{ }
\end{array}
\end{equation}
where $Z$ denotes the $\mathcal{N}=2$, $d=4$ \textit{central charge function}%
, and the Arnowitt-Deser-Misner (ADM) mass \cite{ADM} at the BH horizon is
obtained by extremizing $V_{BH}\left( \phi ,\widetilde{\Gamma }\right) $
with respect to its dependence on the moduli:
\begin{equation}
M_{ADM,H}\left( \widetilde{\Gamma }\right) =\sqrt{\left. V_{BH}\left( \phi ,%
\widetilde{\Gamma }\right) \right| _{\partial _{\phi }V_{BH}=0}}.
\end{equation}

The charge-dependent BH entropy $S_{BH}$ is given by the Bekenstein-Hawking
entropy-area formula \cite{BH1,FGK}
\begin{equation}
S_{BH}\left( \widetilde{\Gamma }\right) =\pi M_{ADM,H}^{2}\left( \widetilde{%
\Gamma }\right) =\frac{A_{H}\left( \widetilde{\Gamma }\right) }{4}=\pi
\left. V_{BH}\left( \phi ,\widetilde{\Gamma }\right) \right| _{\partial
_{\phi }V_{BH}=0}=\pi V_{BH}\left( \phi _{H}\left( \widetilde{\Gamma }%
\right) ,\widetilde{\Gamma }\right) ,  \label{BHEA}
\end{equation}
where $A_{H}$ is the event horizon area. The charge-dependent horizon
configuration $\phi _{H}\left( \widetilde{\Gamma }\right) $ of the real
scalars is obtained by extremizing $V_{BH}\left( \phi ,\widetilde{\Gamma }%
\right) $, \textit{i.e.} by solving the criticality conditions
\begin{equation}
\partial _{\phi }V_{BH}\left( \phi ,\widetilde{\Gamma }\right) =0.
\label{crit-cond-gen}
\end{equation}
Strictly speaking, $\phi _{H}\left( \widetilde{\Gamma }\right) $is an
\textit{attractor} if the critical $\left( 2n_{V}+2\right) \times \left(
2n_{V}+2\right) $ real symmetric Hessian matrix
\begin{equation}
\left. \frac{\partial ^{2}V_{BH}\left( \phi ,\widetilde{\Gamma }\right) }{%
\partial \phi \partial \phi }\right| _{\phi =\phi _{H}\left( \widetilde{%
\Gamma }\right) }  \label{crit-Hessian-gen}
\end{equation}
is a strictly positive-definite matrix\footnote{%
It is worth pointing out that the opposite is in general not true, \textit{%
i.e.} there can be attractor points corresponding to critical Hessian
matrices with some ``flat'' directions (\textit{i.e.} vanishing
eigenvalues). In general, in such a case one has to look at higher-order
covariant derivatives of $V_{BH}$ evaluated at the considered point, and
study their sign. Dependingly on the configurations of the supporting BH
charges, one can obtains stable or unstable critical points. Examples in \
literature of investigations beyond the Hessian level can be found in \cite
{TT,K3,Misra1}.}.

Although non-supersymmetric (non-BPS) BH attractors arise also in $\mathcal{N%
}>2$, $d=4$ and $d=5$ supergravities \cite{FG,FKlast} (see \cite{ADFT} for a
recent review), the richest casistics pertains to $\mathcal{N}=2$, $d=4$
MESGTs, where the manifold parameterized by the scalars is endowed with a
special K\"{a}hler (SK) metric structure.\smallskip

The plan of the paper is as follows.

In Sect. \ref{SKG-gen} we sketchily recall the fundamentals of the local SK
geometry. Thence, in Sect. \ref{Sect2} we introduce the effective BH
potential for a generic $\mathcal{N}=2$, $d=4$ MESGT, and consider its $%
\frac{1}{2}$-BPS critical points \cite{FKS}- \nocite{Strom,FK1,FK2}\cite{FGK}%
, which turn out to be always stable, and thus attractors in strict sense.
Sects. \ref{Sect3} and \ref{Sect4} are devoted to the discussion of the
non-BPS, $Z\neq 0$ case, with an explicit application to the one-modulus
case $n_{V}=1$; in particular, in \ref{Sect3} non-BPS, $Z\neq 0$ critical
points of $V_{BH}$ and the related eigenvalue problem are presented, whereas
Sect. \ref{Sect4} deals with the issue of stability of such a class of
points. Finally, Sect. \ref{Conclusion} contains some summarizing
observations and general remarks, as well as an outlook of possible future
further developments along the considered research directions.
\setcounter{equation}0

\section{\label{SKG-gen}Special K\"{a}hler Geometry}

In the present Section we briefly recall the fundamentals of the SK geometry
underlying the scalar manifold $\mathcal{M}_{n_{V}}$ of $\mathcal{N}=2$, $%
d=4 $ MESGT, $n_{V}$ being the number of Abelian vector supermultiplets
coupled to the supergravity multiplet ($dim_{\mathbb{C}}\mathcal{M}%
_{n_{V}}=n_{V}$).

It is convenient to switch from the Riemannian $2n_{V}$-dim.
parameterization of $\mathcal{M}_{n_{V}}$ given by the local real
coordinates $\left\{ \phi ^{a}\right\} _{a=1,...,2n_{V}}$ to the K\"{a}hler $%
n_{V}$-dim. holomorphic/antiholomorphic parameterization given by the local
complex coordinates $\left\{ z^{i},\overline{z}^{\overline{i}}\right\} _{i,%
\overline{i}=1,...,n_{V}}$. This corresponds to the following \textit{%
unitary Cayley transformation}:
\begin{equation}
z^{k}\equiv \frac{\varphi ^{2k-1}+i\varphi ^{2k}}{\sqrt{2}},~~k=1,...,n_{V}.
\label{unit-transf}
\end{equation}

The metric structure of $\mathcal{M}_{n_{V}}$ is given by the covariant
(special) K\"{a}hler metric tensor $g_{i\overline{j}}\left( z,\overline{z}%
\right) =\partial _{i}\overline{\partial }_{\overline{j}}K\left( z,\overline{%
z}\right) $, $K\left( z,\overline{z}\right) $ being the real K\"{a}hler
potential.

Usually, the $n_{V}\times n_{V}$ Hermitian matrix $g_{i\overline{j}}$ is
assumed to be non-degenerate (\textit{i.e.} invertible, with non-vanishing
determinant and rank $n_{V}$) and with strict positive Euclidean signature (%
\textit{i.e.} with all strictly positive eigenvalues)\textit{\ globally} in $%
\mathcal{M}_{n_{V}}$. We will so assume, even though we will be concerned
mainly with the properties of $g_{i\overline{j}}$ at those peculiar points
of $\mathcal{M}_{n_{V}}$ which are critical points of $V_{BH}$.

It is worth here remarking that various possibilities arise when going
beyond the assumption of \textit{global strict regular }$g_{i\overline{j}}$,
namely:

- (\textit{locally}) \textit{not strictly regular} $g_{i\overline{j}}$,
\textit{i.e. }a (\textit{locally}) non-invertible metric tensor, with some
strictly positive and some vanishing eigenvalues (rank $<n_{V}$);

- (\textit{locally}) \textit{non-regular non-degenerate} $g_{i\overline{j}}$%
, \textit{i.e. }a (\textit{locally}) invertible metric tensor with \textit{%
pseudo-Euclidean signature}, namely with some strictly positive and some
strictly negative eigenvalues (rank $=n_{V}$);

- (\textit{locally}) \textit{non-regular degenerate} $g_{i\overline{j}}$,
\textit{i.e. }a (\textit{locally}) non-invertible metric tensor with some
strictly positive, some strictly negative, and some vanishing eigenvalues
(rank $<n_{V}$).

The \textit{local} violation of strict regularity of $g_{i\overline{j}}$
would produce some kind of ``phase transition'' in the SKG endowing $%
\mathcal{M}_{n_{V}}$, corresponding to a breakdown of the $1$-dim. effective
Lagrangian picture (see \cite{FGK}, \cite{2}, and also \cite{AoB-book} and
\cite{ADFT}) of $d=4$ (extremal) BHs obtained by integrating all massive
states of the theory out, unless new massless states appear \cite{FGK}.

The previously mentioned $\mathcal{N}=2$, $d=4$ \textit{central charge
function} is defined as
\begin{equation}
\begin{array}{l}
Z\left( z,\overline{z};q,p\right) \equiv \widetilde{\Gamma }\Omega V\left( z,%
\overline{z}\right) =q_{\Lambda }L^{\Lambda }\left( z,\overline{z}\right)
-p^{\Lambda }M_{\Lambda }\left( z,\overline{z}\right) =e^{\frac{1}{2}K\left(
z,\overline{z}\right) }\widetilde{\Gamma }\Omega \Pi \left( z\right) = \\
\\
=e^{\frac{1}{2}K\left( z,\overline{z}\right) }\left[ q_{\Lambda }X^{\Lambda
}\left( z\right) -p^{\Lambda }F_{\Lambda }\left( z\right) \right] \equiv e^{%
\frac{1}{2}K\left( z,\overline{z}\right) }W\left( z;q,p\right) ,
\end{array}
\label{Z}
\end{equation}
where $\Omega $ is the $\left( 2n_{V}+2\right) $-dim. square symplectic
metric (subscripts denote dimensions of square sub-blocks)
\begin{equation}
\Omega \equiv \left(
\begin{array}{ccc}
0_{n_{V}+1} &  & -\mathbb{I}_{n_{V}+1} \\
&  &  \\
\mathbb{I}_{n_{V}+1} &  & 0_{n_{V}+1}
\end{array}
\right) ,  \label{Omega}
\end{equation}
and $V\left( z,\overline{z}\right) $ and $\Pi \left( z\right) $ respectively
stand for the $\left( 2n_{V}+2\right) \times 1$ covariantly holomorphic (K%
\"{a}hler weights $\left( 1,-1\right) $) and holomorphic (K\"{a}hler weights
$\left( 2,0\right) $) period vectors in symplectic basis:
\begin{equation}
\begin{array}{c}
\overline{D}\overline{_{i}}V\left( z,\overline{z}\right) =\left( \overline{%
\partial }_{\overline{i}}-\frac{1}{2}\overline{\partial }_{\overline{i}%
}K\right) V\left( z,\overline{z}\right) =0,~~~D_{i}V\left( z,\overline{z}%
\right) =\left( \partial _{i}+\frac{1}{2}\partial _{i}K\right) V\left( z,%
\overline{z}\right)  \\
\Updownarrow  \\
V\left( z,\overline{z}\right) =e^{\frac{1}{2}K\left( z,\overline{z}\right)
}\Pi \left( z\right) ,~~\overline{D}\overline{_{i}}\Pi \left( z\right) =%
\overline{\partial }_{\overline{i}}\Pi \left( z\right) =0,~~~D_{i}\Pi \left(
z\right) =\left( \partial _{i}+\partial _{i}K\right) \Pi \left( z\right) ,
\\
\\
\Pi \left( z\right) \equiv \left(
\begin{array}{c}
X^{\Lambda }\left( z\right)  \\
\\
F_{\Lambda }\left( X\left( z\right) \right)
\end{array}
\right) =exp\left( -\frac{1}{2}K\left( z,\overline{z}\right) \right) \left(
\begin{array}{c}
L^{\Lambda }\left( z,\overline{z}\right)  \\
\\
M_{\Lambda }\left( z,\overline{z}\right)
\end{array}
\right) ,
\end{array}
\label{PI}
\end{equation}
with $X^{\Lambda }\left( z\right) $ and $F_{\Lambda }\left( X\left( z\right)
\right) $ being the holomorphic sections of the $U(1)$ line (Hodge) bundle
over $\mathcal{M}_{n_{V}}$. $W\left( z;q,p\right) $ is the so-called \textit{%
holomorphic }$\mathcal{N}=2$\textit{\ central charge function}, also named $%
\mathcal{N}=2$ \textit{superpotential}. Up to some particular choices of
local symplectic coordinates in $\mathcal{M}_{n_{V}}$, the covariant
symplectic holomorphic sections $F_{\Lambda }\left( X\left( z\right) \right)
$ may be seen as derivatives of an \textit{holomorphic prepotential}
function $F$ (with K\"{a}hler weights $\left( 4,0\right) $):
\begin{equation}
F_{\Lambda }\left( X\left( z\right) \right) =\frac{\partial F\left( X\left(
z\right) \right) }{\partial X^{\Lambda }}.  \label{prepotential}
\end{equation}
In $\mathcal{N}=2$, $d=4$ MESGT the holomorphic function $F$ is constrained
to be homogeneous of degree $2$ in the contravariant symplectic holomorphic
sections $X^{\Lambda }\left( z\right) $, \textit{i.e.} (see \cite{4} and
Refs. therein)
\begin{equation}
2F\left( X\left( z\right) \right) =X^{\Lambda }\left( z\right) F_{\Lambda
}\left( X\left( z\right) \right) .  \label{hom-prop-F}
\end{equation}

The normalization of the holomorphic period vector $\Pi \left( z\right) $ is
such that
\begin{equation}
K\left( z,\overline{z}\right) =-ln\left[ i\left\langle \Pi \left( z\right) ,%
\overline{\Pi }\left( \overline{z}\right) \right\rangle \right] \equiv -ln%
\left[ i\Pi ^{T}\left( z\right) \Omega \overline{\Pi }\left( \overline{z}%
\right) \right] =-ln\left\{ i\left[ \overline{X}^{\Lambda }\left( \overline{z%
}\right) F_{\Lambda }\left( z\right) -X^{\Lambda }\left( z\right) \overline{F%
}_{\Lambda }\left( \overline{z}\right) \right] \right\} ,  \label{norm-PI}
\end{equation}
where $\left\langle \cdot ,\cdot \right\rangle $ stands for the symplectic
scalar product defined by $\Omega $. Note that under a K\"{a}hler
transformation $K\left( z,\overline{z}\right) \longrightarrow K\left( z,%
\overline{z}\right) +f\left( z\right) +\overline{f}\left( \overline{z}%
\right) $ ($f\left( z\right) $ being a generic holomorphic function), the
holomorphic period vector transforms as $\Pi \left( z\right) \longrightarrow
\Pi \left( z\right) e^{-f\left( z\right) }$, and therefore $X^{\Lambda
}\left( z\right) \longrightarrow X^{\Lambda }\left( z\right) e^{-f\left(
z\right) }$. This means that, at least locally, the contravariant
holomorphic symplectic sections $X^{\Lambda }\left( z\right) $ can be
regarded as a set of homogeneous coordinates on $\mathcal{M}_{n_{V}}$,
provided that the Jacobian complex $n_{V}\times n_{V}$ holomorphic matrix
\begin{equation}
e_{i}^{a}\left( z\right) \equiv \frac{\partial }{\partial z^{i}}\left( \frac{%
X^{a}\left( z\right) }{X^{0}\left( z\right) }\right) ,\text{ ~}a=1,...,n_{V}
\end{equation}
is invertible. If this is the case, then one can introduce the local
projective symplectic coordinates
\begin{equation}
t^{a}\left( z\right) \equiv \frac{X^{a}\left( z\right) }{X^{0}\left(
z\right) },
\end{equation}
and the SKG of $\mathcal{M}_{n_{V}}$ turns out to be based on the
holomorphic prepotential $\mathcal{F}\left( t\right) \equiv \left(
X^{0}\right) ^{-2}F\left( X\right) $. By using the $t$-coordinates, Eq. (\ref
{norm-PI}) can be rewritten as follows ($\mathcal{F}_{a}\left( t\right)
=\partial _{a}\mathcal{F}\left( t\right) $, $\overline{t}^{a}=\overline{t^{a}%
}$, $\overline{\mathcal{F}}_{a}\left( \overline{t}\right) =\overline{%
\mathcal{F}_{a}\left( t\right) }$):
\begin{equation}
K\left( t,\overline{t}\right) =-ln\left\{ i\left| X^{0}\left( z\left(
t\right) \right) \right| ^{2}\left[ 2\left( \mathcal{F}\left( t\right) -%
\overline{\mathcal{F}}\left( \overline{t}\right) \right) -\left( t^{a}-%
\overline{t}^{a}\right) \left( \mathcal{F}_{a}\left( t\right) +\overline{%
\mathcal{F}}_{a}\left( \overline{t}\right) \right) \right] \right\} .
\end{equation}
By performing a K\"{a}hler gauge-fixing with $f\left( z\right) =ln\left(
X^{0}\left( z\right) \right) $, yielding that $X^{0}\left( z\right)
\longrightarrow 1$, one thus gets
\begin{equation}
\left. K\left( t,\overline{t}\right) \right| _{X^{0}\left( z\right)
\longrightarrow 1}=-ln\left\{ i\left[ 2\left( \mathcal{F}\left( t\right) -%
\overline{\mathcal{F}}\left( \overline{t}\right) \right) -\left( t^{a}-%
\overline{t}^{a}\right) \left( \mathcal{F}_{a}\left( t\right) +\overline{%
\mathcal{F}}_{a}\left( \overline{t}\right) \right) \right] \right\} .
\label{K-t-X0=1}
\end{equation}
In particular, one can choose the so-called \textit{special coordinates},
i.e. the system of local projective $t$-coordinates such that
\begin{equation}
e_{i}^{a}\left( z\right) =\delta _{i}^{a}\Leftrightarrow t^{a}\left(
z\right) =z^{i}\left( +c^{i},\text{~}c^{i}\in \mathbb{C}\right) .
\end{equation}
Thus, Eq. (\ref{K-t-X0=1}) acquires the form
\begin{equation}
\left. K\left( t,\overline{t}\right) \right| _{X^{0}\left( z\right)
\longrightarrow 1,e_{i}^{a}\left( z\right) =\delta _{i}^{a}}=-ln\left\{ i%
\left[ 2\left( \mathcal{F}\left( z\right) -\overline{\mathcal{F}}\left(
\overline{z}\right) \right) -\left( z^{j}-\overline{z}^{\overline{j}}\right)
\left( \mathcal{F}_{j}\left( z\right) +\overline{\mathcal{F}}_{\overline{j}%
}\left( \overline{z}\right) \right) \right] \right\} .
\end{equation}

Moreover, it should be recalled that $Z$ has K\"{a}hler weights $\left( p,%
\overline{p}\right) =\left( 1,-1\right) $, and therefore its
K\"{a}hler-covariant derivatives read
\begin{equation}
D_{i}Z=\left( \partial _{i}+\frac{1}{2}\partial _{i}K\right) Z,~~\overline{D}%
_{\overline{i}}Z=\left( \overline{\partial }_{\overline{i}}-\frac{1}{2}%
\overline{\partial }_{\overline{i}}K\right) Z.  \label{DiZ}
\end{equation}

The fundamental differential relations of SK geometry are\footnote{%
Actually, there are different (equivalent) defining approaches to SK
geometry. For subtleties and further elucidation concerning such an issue,
see \textit{e.g.} \cite{Craps1} and \cite{Craps2}.} (see \textit{e.g.} \cite
{4}):
\begin{equation}
\left\{
\begin{array}{l}
D_{i}Z=Z_{i}; \\
\\
D_{i}Z_{j}=iC_{ijk}g^{k\overline{k}}\overline{D}_{\overline{k}}\overline{Z}%
=iC_{ijk}g^{k\overline{k}}\overline{Z}_{\overline{k}}; \\
\\
D_{i}\overline{D}_{\overline{j}}\overline{Z}=D_{i}\overline{Z}_{\overline{j}%
}=g_{i\overline{j}}\overline{Z}; \\
\\
D_{i}\overline{Z}=0,
\end{array}
\right.   \label{SKG-rels1}
\end{equation}
where the first relation is nothing but the definition of the so-called
\textit{matter charges} $Z_{i}$, and the fourth relation expresses the K\"{a}%
hler-covariant holomorphicity of $Z$. $C_{ijk}$ is the rank-3, completely
symmetric, covariantly holomorphic tensor of SK geometry (with K\"{a}hler
weights $\left( 2,-2\right) $) (see \textit{e.g.}\footnote{%
Notice that the third of Eqs. (\ref{C}) correctly defines the Riemann tensor
$R_{i\overline{j}k\overline{l}}$, and it is actual the opposite of the one
which may be found in a large part of existing literature. Such a
formulation of the so-called \textit{SKG constraints} is well defined,
because, as we will mention at the end of Sect. \ref{Sect4}, it yields
negative values of the constant scalar curvature of ($n_{V}=1$-dim.)
homogeneous symmetric compact SK manifolds.} \cite{4,Castellani1,DFF}):
\begin{equation}
\begin{array}{l}
\left\{
\begin{array}{l}
C_{ijk}=\left\langle D_{i}D_{j}V,D_{k}V\right\rangle =e^{K}\left( \partial
_{i}\mathcal{N}_{\Lambda \Sigma }\right) D_{j}X^{\Lambda }D_{k}X^{\Sigma }=
\\
\\
=e^{K}\left( \partial _{i}X^{\Lambda }\right) \left( \partial _{j}X^{\Sigma
}\right) \left( \partial _{k}X^{\Xi }\right) \partial _{\Xi }\partial
_{\Sigma }F_{\Lambda }\left( X\right) \equiv e^{K}W_{ijk},\text{~~}\overline{%
\partial }_{\overline{l}}W_{ijk}=0; \\
\\
C_{ijk}=D_{i}D_{j}D_{k}\mathcal{S},~~\mathcal{S}\equiv -iL^{\Lambda
}L^{\Sigma }Im\left( F_{\Lambda \Sigma }\right) ,~~F_{\Lambda \Sigma }\equiv
\frac{\partial F_{\Lambda }}{\partial X^{\Sigma }},F_{\Lambda \Sigma }\equiv
F_{\left( \Lambda \Sigma \right) }~; \\
\\
C_{ijk}=-ig_{i\overline{l}}\overline{f}_{\Lambda }^{\overline{l}%
}D_{j}D_{k}L^{\Lambda },~~~\overline{f}_{\Lambda }^{\overline{l}}\left(
\overline{D}\overline{L}_{\overline{s}}^{\Lambda }\right) \equiv \delta _{%
\overline{s}}^{\overline{l}};
\end{array}
\right.  \\
\\
\overline{D}_{\overline{i}}C_{jkl}=0\text{ (\textit{covariant holomorphicity}%
)}; \\
\\
R_{i\overline{j}k\overline{l}}=-g_{i\overline{j}}g_{k\overline{l}}-g_{i%
\overline{l}}g_{k\overline{j}}+C_{ikp}\overline{C}_{\overline{j}\overline{l}%
\overline{p}}g^{p\overline{p}}\text{ (usually named \textit{SKG constraints})%
}; \\
\\
D_{[i}C_{j]kl}=0,
\end{array}
\label{C}
\end{equation}
where the last property is a consequence, through the SKG constraints and
the covariant holomorphicity of $C_{ijk}$, of the Bianchi identities for the
Riemann tensor $R_{i\overline{j}k\overline{l}}$, and square brackets denote
antisymmetrization with respect to enclosed indices. It is worth recalling
that in a generic K\"{a}hler geometry $R_{i\overline{j}k\overline{l}}$ reads
\begin{equation}
\begin{array}{l}
R_{i\overline{j}k\overline{l}}=g^{m\overline{n}}\left( \overline{\partial }_{%
\overline{l}}\overline{\partial }_{\overline{j}}\partial _{m}K\right)
\partial _{i}\overline{\partial }_{\overline{n}}\partial _{k}K-\overline{%
\partial }_{\overline{l}}\partial _{i}\overline{\partial }_{\overline{j}%
}\partial _{k}K=g_{k\overline{n}}\partial _{i}\overline{\Gamma }_{\overline{l%
}\overline{j}}^{~~\overline{n}}=g_{n\overline{l}}\overline{\partial }_{%
\overline{j}}\Gamma _{ki}^{~~n}, \\
\\
\overline{R_{i\overline{j}k\overline{l}}}=R_{j\overline{i}l\overline{k}}%
\text{ \ \ \ (\textit{reality})}, \\
\\
\Gamma _{ij}^{~~l}=-g^{l\overline{l}}\partial _{i}g_{j\overline{l}}=-g^{l%
\overline{l}}\partial _{i}\overline{\partial }_{\overline{l}}\partial
_{j}K=\Gamma _{\left( ij\right) }^{~~l},
\end{array}
\text{\ }
\end{equation}
where $\Gamma _{ij}^{~~l}$ stand for the Christoffel symbols of the second
kind of the K\"{a}hler metric $g_{i\overline{j}}$.

In the first of Eqs. (\ref{C}), a fundamental entity, the so-called kinetic
matrix $\mathcal{N}_{\Lambda \Sigma }\left( z,\overline{z}\right) $ of $%
\mathcal{N}=2$, $d=4$ MESGT, has been introduced. It is an $\left(
n_{V}+1\right) \times \left( n_{V}+1\right) $ complex symmetric,
moduli-dependent, K\"{a}hler gauge-invariant matrix defined by the following
fundamental \textit{Ans\"{a}tze} of SKG, solving the \textit{SKG constraints}
(given by the third of Eqs. (\ref{C})):
\begin{equation}
M_{\Lambda }=\mathcal{N}_{\Lambda \Sigma }L^{\Sigma },~~D_{i}M_{\Lambda }=%
\overline{\mathcal{N}}_{\Lambda \Sigma }D_{i}L^{\Sigma }.  \label{Ans1}
\end{equation}
By introducing the $\left( n_{V}+1\right) \times \left( n_{V}+1\right) $
complex matrices ($I=1,...,n_{V}+1$)
\begin{equation}
f_{I}^{\Lambda }\left( z,\overline{z}\right) \equiv \left( \overline{D}_{%
\overline{i}}\overline{L}^{\Lambda }\left( z,\overline{z}\right) ,L^{\Lambda
}\left( z,\overline{z}\right) \right) ,\text{ \ }h_{I\Lambda }\left( z,%
\overline{z}\right) \equiv \left( \overline{D}_{\overline{i}}\overline{M}%
_{\Lambda }\left( z,\overline{z}\right) ,M_{\Lambda }\left( z,\overline{z}%
\right) \right) ,
\end{equation}
the \textit{Ans\"{a}tze} (\ref{Ans1}) univoquely determine $\mathcal{N}%
_{\Lambda \Sigma }\left( z,\overline{z}\right) $ as
\begin{equation}
\mathcal{N}_{\Lambda \Sigma }\left( z,\overline{z}\right) =h_{I\Lambda
}\left( z,\overline{z}\right) \circ \left( f^{-1}\right) _{\Sigma
}^{I}\left( z,\overline{z}\right) ,
\end{equation}
where $\circ $ denotes the usual matrix product, and $\left( f^{-1}\right)
_{\Sigma }^{I}f_{I}^{\Lambda }=\delta _{\Sigma }^{\Lambda }$, $\left(
f^{-1}\right) _{\Lambda }^{I}f_{J}^{\Lambda }=\delta _{J}^{I}$.

The covariantly holomorphic $\left( 2n_{V}+2\right) \times 1$ period vector $%
V\left( z,\overline{z}\right) $ is \textit{symplectically orthogonal} to all
its K\"{a}hler-covariant derivatives:
\begin{equation}
\left\{
\begin{array}{l}
\left\langle V\left( z,\overline{z}\right) ,D_{i}V\left( z,\overline{z}%
\right) \right\rangle =0; \\
\\
\left\langle V\left( z,\overline{z}\right) ,\overline{D}_{\overline{i}%
}V\left( z,\overline{z}\right) \right\rangle =0; \\
\\
\left\langle V\left( z,\overline{z}\right) ,D_{i}\overline{V}\left( z,%
\overline{z}\right) \right\rangle =0; \\
\\
\left\langle V\left( z,\overline{z}\right) ,\overline{D}_{\overline{i}}%
\overline{V}\left( z,\overline{z}\right) \right\rangle =0.
\end{array}
\right.  \label{ortho-rels}
\end{equation}
Morover, it holds that
\begin{equation}
\begin{array}{l}
g_{i\overline{j}}\left( z,\overline{z}\right) =-i\left\langle D_{i}V\left( z,%
\overline{z}\right) ,\overline{D}_{\overline{j}}\overline{V}\left( z,%
\overline{z}\right) \right\rangle = \\
\\
=-2Im\left( \mathcal{N}_{\Lambda \Sigma }\left( z,\overline{z}\right)
\right) D_{i}L^{\Lambda }\left( z,\overline{z}\right) \overline{D}_{%
\overline{i}}\overline{L}^{\Sigma }\left( z,\overline{z}\right) =2Im\left(
F_{\Lambda \Sigma }\left( z\right) \right) D_{i}L^{\Lambda }\left( z,%
\overline{z}\right) \overline{D}_{\overline{i}}\overline{L}^{\Sigma }\left(
z,\overline{z}\right) ;
\end{array}
\label{ortho1}
\end{equation}
\begin{equation}
\left\langle V\left( z,\overline{z}\right) ,D_{i}\overline{D}_{\overline{j}%
}V\left( z,\overline{z}\right) \right\rangle =iC_{ijk}g^{k\overline{k}%
}\left\langle V\left( z,\overline{z}\right) ,\overline{D}_{\overline{k}}%
\overline{V}\left( z,\overline{z}\right) \right\rangle =0.  \label{ortho2}
\end{equation}

The fundamental $\left( 2n_{V}+2\right) \times 1$ vector identity defining
the geometric structure of SK manifolds read as follows \cite
{FBC,K1,K2,BFM,AoB-book,K3}:
\begin{equation}
\widetilde{\Gamma }^{T}-i\Omega \mathcal{M}\left( \mathcal{N}\right)
\widetilde{\Gamma }^{T}=-2iZ\overline{V}-2ig^{j\overline{j}}\left( \overline{%
D}_{\overline{j}}\overline{Z}\right) D_{j}V.  \label{SKG-identities1}
\end{equation}
The $\left( 2n_{V}+2\right) \times \left( 2n_{V}+2\right) $ real symmetric
matrix $\mathcal{M}\left( \mathcal{N}\right) $ is defined as \cite{4,FK1,FK2}
\begin{eqnarray}
\mathcal{M}\left( \mathcal{N}\right) &=&\mathcal{M}\left( Re\left( \mathcal{N%
}\right) ,Im\left( \mathcal{N}\right) \right) \equiv \\
&\equiv &\left(
\begin{array}{ccc}
Im\left( \mathcal{N}\right) +Re\left( \mathcal{N}\right) \left( Im\left(
\mathcal{N}\right) \right) ^{-1}Re\left( \mathcal{N}\right) &  & -Re\left(
\mathcal{N}\right) \left( Im\left( \mathcal{N}\right) \right) ^{-1} \\
-\left( Im\left( \mathcal{N}\right) \right) ^{-1}Re\left( \mathcal{N}\right)
&  & \left( Im\left( \mathcal{N}\right) \right) ^{-1}
\end{array}
\right) ,  \notag
\end{eqnarray}
where $\mathcal{N}_{\Lambda \Sigma }$ is a complex symmetric matrix playing
a key role in $\mathcal{N}=2$, $d=4$ MESGT (see \textit{e.g.} the report
\cite{4}). It is worth reminding that $\mathcal{M}\left( \mathcal{N}\right) $
is symplectic with respect to the metric $\Omega $ defined in Eq. (\ref
{Omega}), \textit{i.e.} it satisfies ($\left( \mathcal{M}\left( \mathcal{N}%
\right) \right) ^{T}=\mathcal{M}\left( \mathcal{N}\right) $)
\begin{equation}
\mathcal{M}\left( \mathcal{N}\right) \Omega \mathcal{M}\left( \mathcal{N}%
\right) =\Omega .
\end{equation}

By using Eqs. (\ref{norm-PI}), (\ref{ortho-rels}), (\ref{ortho1}) and (\ref
{ortho2}), the identity (\ref{SKG-identities1}) implies the following
relations:
\begin{equation}
\left\{
\begin{array}{l}
\left\langle V,\widetilde{\Gamma }^{T}-i\Omega \mathcal{M}\left( \mathcal{N}%
\right) \widetilde{\Gamma }^{T}\right\rangle =-2Z; \\
\\
\left\langle \overline{V},\widetilde{\Gamma }^{T}-i\Omega \mathcal{M}\left(
\mathcal{N}\right) \widetilde{\Gamma }^{T}\right\rangle =0; \\
\\
\left\langle D_{i}V,\widetilde{\Gamma }^{T}-i\Omega \mathcal{M}\left(
\mathcal{N}\right) \widetilde{\Gamma }^{T}\right\rangle =0; \\
\\
\left\langle \overline{D}_{\overline{i}}\overline{V},\widetilde{\Gamma }%
^{T}-i\Omega \mathcal{M}\left( \mathcal{N}\right) \widetilde{\Gamma }%
^{T}\right\rangle =-2\overline{D}_{\overline{i}}\overline{Z}.
\end{array}
\right.  \label{SKG-SKG-yielded}
\end{equation}

There are only $2n_{V}$ independent real relations out of the $4n_{V}+4$
real ones yielded by the $2n_{V}+2$ complex identities (\ref{SKG-identities1}%
). Indeed, by taking the real and imaginary part of the SKG vector identity (%
\ref{SKG-identities1}) one respectively obtains
\begin{equation}
\widetilde{\Gamma }^{T}=-2Re\left[ iZ\overline{V}+iG^{j\overline{j}}\left(
\overline{D}_{\overline{j}}\overline{Z}\right) D_{j}V\right] =-2Im\left[
\overline{Z}V+G^{j\overline{j}}\left( D_{j}Z\right) \left( \overline{D}_{%
\overline{j}}\overline{V}\right) \right] ;  \label{Re-SKG-SKG}
\end{equation}
\begin{equation}
\Omega \mathcal{M}\left( \mathcal{N}\right) \widetilde{\Gamma }^{T}=2Im\left[
iZ\overline{V}+iG^{j\overline{j}}\left( \overline{D}_{\overline{j}}\overline{%
Z}\right) D_{j}V\right] =2Re\left[ \overline{Z}V+G^{j\overline{j}}\left(
D_{j}Z\right) \left( \overline{D}_{\overline{j}}\overline{V}\right) \right] .
\label{Im-SKG-SKG}
\end{equation}
Consequently, the imaginary and real parts of the SKG vector identity (\ref
{SKG-identities1}) are \textit{linearly dependent} one from the other, being
related by the $\left( 2n_{V}+2\right) \times \left( 2n_{V}+2\right) $ real
matrix
\begin{equation}
\Omega \mathcal{M}\left( \mathcal{N}\right) =\left(
\begin{array}{ccc}
\left( Im\left( \mathcal{N}\right) \right) ^{-1}Re\left( \mathcal{N}\right)
&  & -\left( Im\left( \mathcal{N}\right) \right) ^{-1} \\
Im\left( \mathcal{N}\right) +Re\left( \mathcal{N}\right) \left( Im\left(
\mathcal{N}\right) \right) ^{-1}Re\left( \mathcal{N}\right) &  & -Re\left(
\mathcal{N}\right) \left( Im\left( \mathcal{N}\right) \right) ^{-1}
\end{array}
\right) .  \label{epsilon-emme}
\end{equation}
Put another way, Eqs. (\ref{Re-SKG-SKG}) and (\ref{Im-SKG-SKG}) yield
\begin{equation}
Re\left[ Z\overline{V}+G^{j\overline{j}}\left( \overline{D}_{\overline{j}}%
\overline{Z}\right) D_{j}V\right] =\Omega \mathcal{M}\left( \mathcal{N}%
\right) Im\left[ Z\overline{V}+G^{j\overline{j}}\left( \overline{D}_{%
\overline{j}}\overline{Z}\right) D_{j}V\right] ,  \label{rotation1}
\end{equation}
expressing the fact that the real and imaginary parts of the quantity $Z%
\overline{V}+G^{j\overline{j}}\left( \overline{D}_{\overline{j}}\overline{Z}%
\right) D_{j}V$ are simply related through a \textit{symplectic rotation}
given by the matrix $\Omega \mathcal{M}\left( \mathcal{N}\right) $, whose
simplecticity directly follows from the symplectic nature of $\mathcal{M}%
\left( \mathcal{N}\right) $. Eq. (\ref{rotation1}) reduces the number of
independent real relations implied by the identity (\ref{SKG-identities1})
from $4n_{V}+4$ to $2n_{V}+2$.

Moreover, it should be stressed that vector identity (\ref{SKG-identities1})
entails 2 \textit{redundant} degrees of freedom, encoded in the homogeneity
(of degree 1) of (\ref{SKG-identities1}) under complex rescalings of $%
\widetilde{\Gamma }$. Indeed, by using the definition (\ref{Z}), it is easy
to check that the right-hand side of (\ref{SKG-identities1}) gets rescaled
by an overall factor $\lambda $ under the following transformation on $%
\widetilde{\Gamma }$:
\begin{equation}
\widetilde{\Gamma }\longrightarrow \lambda \widetilde{\Gamma },~~~\lambda
\in \mathbb{C}.
\end{equation}
Thus, as announced, only $2n_{V}$ real independent relations are actually
yielded by the vector identity (\ref{SKG-identities1}).

This is clearly consistent with the fact that the $2n_{V}+2$\ complex
identities (\ref{SKG-identities1}) express nothing but a \textit{change of
basis} of the BH charge configurations, between the K\"{a}hler-invariant $%
1\times \left( 2n_{V}+2\right) $\ symplectic (magnetic/electric) basis
vector $\widetilde{\Gamma }$ defined by Eq. (\ref{Gamma-tilde}) and the
complex, moduli-dependent $1\times \left( n_{V}+1\right) $ \textit{matter
charges} vector (with K\"{a}hler weights $\left( 1,-1\right) $)
\begin{equation}
\mathcal{Z}\left( z,\overline{z}\right) \equiv \left( Z\left( z,\overline{z}%
\right) ,Z_{i}\left( z,\overline{z}\right) \right) _{i=1,...,n_{V}}.
\label{Z-call}
\end{equation}

It should be recalled that the BH charges are conserved due to the overall $%
\left( U(1)\right) ^{n_{V}+1}$ gauge-invariance of the system under
consideration, and $\widetilde{\Gamma }$ and $\mathcal{Z}\left( z,\overline{z%
}\right) $ are two \textit{equivalent} basis for them. Their very
equivalence relations are given by the SKG identities (\ref{SKG-identities1}%
) themselves. By its very definition (\ref{Gamma-tilde}), $\widetilde{\Gamma
}$\ is \textit{moduli-independent} (at least in a stationary, spherically
symmetric and asymptotically flat extremal BH background, as it is the case
being treated here), whereas $Z$ is \textit{moduli-dependent}, since it
refers to the eigenstates of the $\mathcal{N}=2$, $d=4$ supergravity
multiplet and of the $n_{V}$\ Maxwell vector supermultiplets.
\setcounter{equation}0

\section{\label{Sect2}Supersymmetric Attractors}

The ``effective BH potential'' of $\mathcal{N}=2$, $d=4$ MESGT has the
following expression \cite{FK1,FK2,4} :
\begin{equation}
V_{BH}\left( z,\overline{z};q,p\right) =\left| Z\right| ^{2}\left( z,%
\overline{z};q,p\right) +g^{j\overline{j}}\left( z,\overline{z}\right)
D_{j}Z\left( z,\overline{z};q,p\right) \overline{D}_{\overline{j}}\overline{Z%
}\left( z,\overline{z};q,p\right) .  \label{VBH1}
\end{equation}
An elegant way to obtain $V_{BH}$ is given by left-multiplying the SKG
vector identity (\ref{SKG-identities1}) by the $1\times \left(
2n_{V}+2\right) $ complex moduli-dependent vector $-\frac{1}{2}\widetilde{%
\Gamma }\mathcal{M}\left( \mathcal{N}\right) $; due to the symplecticity of
the matrix $\mathcal{M}\left( \mathcal{N}\right) $, one obtains \cite
{FK1,FK2,4}
\begin{equation}
V_{BH}\left( z,\overline{z};q,p\right) =-\frac{1}{2}\widetilde{\Gamma }%
\mathcal{M}\left( \mathcal{N}\right) \widetilde{\Gamma }^{T}.  \label{VBH2}
\end{equation}
Thus, $V_{BH}$ is identified with the first (of two), lowest-order
(-quadratic- in charges), positive-definite real invariant $I_{1}$ of SK
geometry (see \textit{e.g.} \cite{K3,4}).\ It is worth noticing that the
result (\ref{VBH2}) can also be derived from the SK geometry identities (\ref
{SKG-identities1}) by using the relation (see \cite{FKlast}, where a
generalization for $\mathcal{N}>2$-extended supergravities is also given)%
\textbf{\ }
\begin{equation}
\frac{1}{2}\left( \mathcal{M}\left( \mathcal{N}\right) +i\Omega \right)
\mathcal{V}=i\Omega \mathcal{V}\Leftrightarrow \mathcal{M}\left( \mathcal{N}%
\right) \mathcal{V}=i\Omega \mathcal{V},
\end{equation}
where $\mathcal{V}$ is a $\left( 2n_{V}+2\right) \times \left(
n_{V}+1\right) $ matrix defined as:
\begin{equation}
\mathcal{V}\equiv \left( V,\overline{D}_{\overline{1}}\overline{V},...,%
\overline{D}_{\overline{n_{V}}}\overline{V}\right) .
\end{equation}

By differentiating Eq. (\ref{VBH1}) with respect to the scalars, it is easy
to check that the general criticality conditions (\ref{crit-cond-gen})
acquire the peculiar form \cite{FGK}
\begin{equation}
D_{i}V_{BH}=\partial _{i}V_{BH}=0\Leftrightarrow 2\overline{Z}D_{i}Z+g^{j%
\overline{j}}\left( D_{i}D_{j}Z\right) \overline{D}_{\overline{j}}\overline{Z%
}=0;  \label{AEs1}
\end{equation}
this is what one should rigorously call the $\mathcal{N}=2$, $d=4$ MESGT
attractor Eqs. (AEs). By means of the features of SKG given by Eqs. (\ref
{SKG-rels1}), the $\mathcal{N}=2$ AEs (\ref{AEs1}) can be re-expressed as
follows \cite{FGK}:
\begin{equation}
2\overline{Z}Z_{i}+iC_{ijk}g^{j\overline{j}}g^{k\overline{k}}\overline{Z}_{%
\overline{j}}\overline{Z}_{\overline{k}}=0.  \label{AEs2}
\end{equation}
It is evident that the tensor $C_{ijk}$ is crucial in relating the $\mathcal{%
N}=2$ central charge function $Z$ (\textit{graviphoton charge}) and the $%
n_{V}$ \textit{matter charges} $Z_{i}$ (coming from the $n_{V}$ Abelian
vector supermultiplets) at the critical points of $V_{BH}$ in the SK scalar
manifold $\mathcal{M}_{n_{V}}$.

The static, spherically symmetric, asymptotically flat BHs are known to be
described by an effective $d=1$ Lagrangian (\cite{FGK}, \cite{2}, and also
\cite{AoB-book} and \cite{ADFT}), with an effective scalar potential and
effective fermionic ``mass terms'' terms controlled by the vector $%
\widetilde{\Gamma }$ of the field-strength fluxes (defined by Eq. (\ref
{Gamma-tilde})). The \textit{``apparent'' gravitino mass} is given by $Z$,
whereas the \textit{gaugino mass matrix} $\Lambda _{ij}$ reads (see the
second Ref. of \cite{DFF})
\begin{equation}
\Lambda _{ij}=C_{ijk}g^{k\overline{k}}\overline{Z}_{\overline{k}}.
\end{equation}
The \textit{supersymmetry breaking order parameters}, related to the mixed
gravitino-gaugino couplings, are nothing but the \textit{matter charge(
function)s} $D_{i}Z=Z_{i}$ (see the first of Eqs. (\ref{SKG-rels1})).

As evident from the AEs (\ref{AEs1}) and (\ref{AEs2}), the conditions
\begin{equation}
(Z\neq 0,)~D_{i}Z=0\text{~~~}\forall i=1,...,n_{V}  \label{BPS-conds}
\end{equation}
determine a (\textit{non-degenerate}) critical point of $V_{BH}$, namely a $%
\frac{1}{2}$-BPS critical point, which preserve 4 supersymmetry degrees of
freedom out of the 8 pertaining to the $\mathcal{N}=2$, $d=4$ Poincar\`{e}
superalgebra related to the asymptotical Minkowski background. The horizon
ADM squared mass at $\frac{1}{2}$-BPS critical points of $V_{BH}$ saturates
the BPS bound, reading \cite{FKS}-\nocite{Strom,FK1,FK2}\cite{FGK}:\textbf{\
}
\begin{equation}
M_{ADM,H,\frac{1}{2}-BPS}^{2}=\left. V_{BH}\right| _{\frac{1}{2}-BPS}=\left|
Z\right| _{\frac{1}{2}-BPS}^{2}+\left[ g^{i\overline{i}}\left( D_{i}Z\right)
\left( \overline{D}_{\overline{i}}\overline{Z}\right) \right] _{\frac{1}{2}%
-BPS}=\left| Z\right| _{\frac{1}{2}-BPS}^{2}>0.  \label{V-BPS}
\end{equation}

In general, $\frac{1}{2}$-BPS critical points are (at least local) minima of
$V_{BH}$ in $\mathcal{M}_{n_{V}}$, and therefore they are stable; thus, they
are \textit{attractors} in strict sense. Indeed, the $2n_{V}\times 2n_{V}$
(covariant) Hessian matrix (in $\left( z,\overline{z}\right) $-coordinates)
of $V_{BH}$ evaluated at such points is strictly positive-definite \cite{FGK}
:
\begin{eqnarray}
&&
\begin{array}{l}
\left( D_{i}D_{j}V_{BH}\right) _{\frac{1}{2}-BPS}=\left( \partial
_{i}\partial _{j}V_{BH}\right) _{\frac{1}{2}-BPS}=0, \\
\\
\left( D_{i}\overline{D}_{\overline{j}}V_{BH}\right) _{\frac{1}{2}%
-BPS}=\left( \partial _{i}\overline{\partial }_{\overline{j}}V_{BH}\right) _{%
\frac{1}{2}-BPS}=2\left( g_{i\overline{j}}V_{BH}\right) _{\frac{1}{2}%
-BPS}=2\left. g_{i\overline{j}}\right| _{\frac{1}{2}-BPS}\left| Z\right| _{%
\frac{1}{2}-BPS}^{2}>0,
\end{array}
\notag \\
&&  \label{SUSY-crit}
\end{eqnarray}
where here and below the notation ``$>0$'' (``$<0$'') is understood as
strict positive-(negative-)definiteness. Eqs. (\ref{SUSY-crit}) yield that
the Hermiticity and (strict) positive-definiteness of the (covariant)
Hessian matrix (in $\left( z,\overline{z}\right) $-coordinates) of $V_{BH}$
at the $\frac{1}{2}$-BPS critical points are due to the Hermiticity and -
assumed - (strict) positive-definiteness (actually holding globally) of the
metric $g_{i\overline{j}}$ of $\mathcal{M}_{n_{V}}$.

Considering the $\mathcal{N}=2$, $d=4$ MESGT Lagrangian in a static,
spherically symmetric, asymptotically flat BH background, and denoting by $%
\psi $ and $\lambda ^{i}$ respectively the gravitino and gaugino fields, it
is easy to see that such a Lagrangian contains terms of the form (see the
second and third Refs. of \cite{DFF})
\begin{equation}
\begin{array}{l}
Z\psi \psi ; \\
\\
C_{ijk}g^{k\overline{k}}\left( \overline{D}_{\overline{k}}\overline{Z}%
\right) \lambda ^{i}\lambda ^{j}; \\
\\
\left( D_{i}Z\right) \lambda ^{i}\psi .
\end{array}
\end{equation}
Thus, the ($\frac{1}{2}$)-BPS conditions (\ref{BPS-conds}) implies the
gaugino mass term and the $\lambda \psi $ term to vanish at the $\frac{1}{2}$%
-BPS critical points of $V_{BH}$ in $\mathcal{M}_{n_{V}}$. It is interesting
to remark that the gravitino ``apparent mass'' term $Z\psi \psi $ is in
general non-vanishing, also when evaluated at the considered $\frac{1}{2}$%
-BPS attractors; this is ultimately a consequence of the fact that the
extremal BH horizon geometry at the $\frac{1}{2}$-BPS (as well as at the
non-BPS) attractors is Bertotti-Robinson $AdS_{2}\times S^{2}$ \cite
{BR1,BR2,BR3}. \setcounter{equation}0

\section{\label{Sect3}Non-BPS Critical Points of $V_{BH}$ with $Z\neq 0$}

It is here worth recalling once again that what we call extremal BH \textit{%
attractor} in (asymptotically flat) $\mathcal{N}=2$, $d=4$ MESGT is,
strictly speaking, a configuration of the scalar fluctuations which is a(n
at least local) minimum for the ``effective BH potential'' $V_{BH}$ (as also
pointed out in \cite{GIJT}), seen as a positive-definite, real function in
the SK scalar manifold $\mathcal{M}_{n_{V}}$. Put another way, an extremal
BH attractor (horizon) scalar configuration satisfies the AEs (\ref{AEs1})
or (\ref{AEs2}), and it is furthermore constrained by the condition of
positive-definiteness of the Hessian matrix of $V_{BH}$, shorthand denoted
as
\begin{equation}
\left( \partial _{i}\partial _{j}V_{BH}\right) _{\partial V_{BH}=0}>0.
\label{pos}
\end{equation}
Obviously, the $\frac{1}{2}$-BPS conditions (\ref{BPS-conds}) are not the
most general ones satisfying the AEs (\ref{AEs1}) or (\ref{AEs2}). For
instance, one might consider critical points of $V_{BH}$ (thus satisfying
the AEs (\ref{AEs1}) or (\ref{AEs2})) characterized by
\begin{equation}
\left\{
\begin{array}{l}
D_{i}Z\neq 0,\text{ for \textit{(at least one)} }i, \\
Z\neq 0.
\end{array}
\right.  \label{non-BPS-Z<>0}
\end{equation}
Such critical points are \textit{non-supersymmetric} ones (\textit{i.e.}
they do \textit{not} preserve any of the 8 supersymmetry degrees of freedom
of the asymptotical Minkowski background), and they correspond to an
extremal, non-BPS BH background. They are commonly named \textit{non-BPS} $%
Z\neq 0$ \textit{critical points of }$V_{BH}$. We will devote the present
Sect. (and, after a general treatment, also next Sect. \ref{Sect4}) to
present their main features.

The horizon ADM squared mass corresponding to non-BPS $Z\neq 0$ critical
points of $V_{BH}$ does \textit{not} saturate the BPS bound (\cite{K1}, \cite
{K2}, \cite{Tom}):
\begin{equation}
\begin{array}{l}
M_{ADM,H,non-BPS,Z\neq 0}^{2}=\left. V_{BH}\right| _{non-BPS,Z\neq 0}= \\
\\
=\left| Z\right| _{non-BPS,Z\neq 0}^{2}+\left[ g^{i\overline{i}}\left(
D_{i}Z\right) \left( \overline{D}_{\overline{i}}\overline{Z}\right) \right]
_{non-BPS,Z\neq 0}>\left| Z\right| _{non-BPS,Z\neq 0}^{2}.
\end{array}
\label{V-non-BPS}
\end{equation}
As implied by AEs (\ref{AEs2}), if at non-BPS $Z\neq 0$ critical points it
holds that $D_{i}Z\neq 0$ for at least one index $i$ and $Z\neq 0$, then
\begin{equation}
\left( C_{ijk}\right) _{non-BPS,Z\neq 0}\neq 0,~~~\text{for some }\left(
i,j,k\right) \in \left\{ 1,...,n_{V}\right\} ^{3},
\end{equation}
\textit{i.e.} the SKG rank-3 symmetric tensor will for sure have some
non-vanishing components in order for criticality conditions (\ref{AEs2})\
to be satisfied at non-BPS $Z\neq 0$ critical points.

Moreover, the general criticality conditions (\ref{AEs1}) for $V_{BH}$ can
be recognized to be the general Ward identities relating the gravitino mass $%
Z$, the gaugino masses $D_{i}D_{j}Z$ and the supersymmetry-breaking order
parameters $D_{i}Z$ in a generic spontaneously broken supergravity theory
\cite{9}. Indeed, away from $\frac{1}{2}$-BPS critical points (\textit{i.e.}
for $D_{i}Z\neq 0$ for some $i$), the AEs (\ref{AEs1}) can be re-expressed
as follows:
\begin{equation}
\left( \mathbf{M}_{ij}h^{j}\right) _{\partial V_{BH}=0}=0,  \label{EQ}
\end{equation}
with
\begin{equation}
\mathbf{M}_{ij}\equiv D_{i}D_{j}Z+2\frac{\overline{Z}}{\left[ g^{k\overline{k%
}}\left( D_{k}Z\right) \left( \overline{D}_{\overline{k}}\overline{Z}\right) %
\right] }\left( D_{i}Z\right) \left( D_{j}Z\right) ,\text{ }(\text{%
K\"{a}hler weights }\left( 1,-1\right) ),
\end{equation}
and
\begin{equation}
h^{j}\equiv g^{j\overline{j}}\overline{D}_{\overline{j}}\overline{Z},\text{ }%
(\text{K\"{a}hler weights }\left( -1,1\right) ).
\end{equation}

For a non-vanishing contravariant vector $h^{j}$ (\textit{i.e. }away from $%
\frac{1}{2}$-BPS critical points, as pointed out above), Eq. (\ref{EQ})
admits a solution iff the $n_{V}\times n_{V}$ complex symmetric matrix $%
\mathbf{M}_{ij}$ has vanishing determinant (implying that it has at most $%
n_{V}-1$ non-vanishing eigenvalues) at the considered (non-BPS) critical
points of $V_{BH}$ (however, notice that $\mathbf{M}_{ij}$\ is symmetric but
not necessarily Hermitian, thus in general its eigenvalues are not
necessarily real)\smallskip .

$n_{V}=1$ SKG represents a noteworthy case, in which major simplifications
occur.

Indeed, in the one-modulus case the condition of vanishing determinant
trivially reads ($z^{1}\equiv z$)
\begin{equation}
\mathbf{M}_{11}=0,
\end{equation}
and (away from $\frac{1}{2}$-BPS critical points, \textit{i.e.} for $%
D_{z}Z\neq 0$) it is equivalent to the criticality condition $\partial
_{z}V_{BH}=0$. $n_{V}=1$ AEs (\ref{AEs2}) consist of the unique complex Eq.
\begin{equation}
\partial _{z}V_{BH}=0\Leftrightarrow 2\overline{Z}D_{z}Z+iCg^{-2}\left(
\overline{D}_{\overline{z}}\overline{Z}\right) ^{2}=0,
\end{equation}
where we defined $C_{111}\equiv C\left( z,\overline{z}\right) \in \mathbb{C}$
and $g_{1\overline{1}}\equiv g\left( z,\overline{z}\right) \in \mathbb{R}%
_{0}^{+}$. From the treatment given above, it necessarily holds that
\begin{equation}
\left\{
\begin{array}{l}
C_{non-BPS,Z\neq 0}\neq 0, \\
\\
\left| D_{z}Z\right| _{non-BPS,Z\neq 0}^{2}=4\left[ g^{4}\frac{\left|
Z\right| ^{2}}{\left| C\right| ^{2}}\right] _{non-BPS,Z\neq 0}>0.
\end{array}
\right.  \label{rel1}
\end{equation}
Consequently, the horizon ADM squared mass at non-BPS $Z\neq 0$ critical
points of $V_{BH}$ in $n_{V}=1$ SKG reads
\begin{equation}
\begin{array}{l}
M_{ADM,H,non-BPS,Z\neq 0}^{2}=\left. V_{BH}\right| _{non-BPS,Z\neq 0}=\left|
Z\right| _{non-BPS,Z\neq 0}^{2}+g^{-1}\left| D_{z}Z\right| _{non-BPS,Z\neq
0}^{2}= \\
\\
=\left| Z\right| _{non-BPS,Z\neq 0}^{2}\left[ 1+4\left( \frac{g^{3}}{\left|
C\right| ^{2}}\right) _{non-BPS,Z\neq 0}\right] >\left| Z\right|
_{non-BPS,Z\neq 0}^{2}.
\end{array}
\label{V-non-BPS-2}
\end{equation}
Eq. (\ref{V-non-BPS-2}) yields an interesting feature of non-BPS $Z\neq 0$
critical points of $V_{BH}$ in $n_{V}=1$ SKG: the entropy $%
S_{BH,non-BPS,Z\neq 0}=\pi \left. V_{BH}\right| _{non-BPS,Z\neq 0}$ is
multiplicatively (and increasingly) ``renormalized'' (with respect to its
formal expression in the $\frac{1}{2}$-BPS case - see Eq. (\ref{V-BPS}) - )
as follows:
\begin{equation}
S_{BH,non-BPS,Z\neq 0}=\pi \gamma \left| Z\right| _{non-BPS,Z\neq 0}^{2},
\end{equation}
with
\begin{equation}
\gamma -1\equiv 4\left( \frac{g^{3}}{\left| C\right| ^{2}}\right)
_{non-BPS,Z\neq 0}>0.  \label{gamma}
\end{equation}

Now, let us introduce the so-called non-BPS $Z\neq 0$ \textit{scalar
``supersymmetry breaking order parameter''} as
\begin{equation}
\mathcal{O}_{non-BPS,Z\neq 0}\equiv \frac{\left( g^{i\overline{j}}D_{i}Z%
\overline{D}_{\overline{j}}\overline{Z}\right) _{non-BPS,Z\neq 0}}{\left|
Z\right| _{non-BPS,Z\neq 0}^{2}}>0,  \label{gamma3}
\end{equation}
the strict positivity bound directly coming from the assumed (global) strict
positive definiteness of the metric $g_{i\overline{j}}$ of $\mathcal{M}%
_{n_{V}}$. The actual independence of $\mathcal{O}_{non-BPS,Z\neq 0}$ on $%
\left| Z\right| _{non-BPS,Z\neq 0}^{2}$ determines the multiplicative (and
increasing) ``renormalization'' of $S_{BH,non-BPS,Z\neq 0}$ to occur.
Nevertheless, the definition (\ref{gamma3}) clearly holds $\forall n\in
\mathbb{N}$, also when no multiplicative ``renormalization'' takes place.

It is immediate to conclude that $\gamma -1$ can be identified with $%
\mathcal{O}_{non-BPS,Z\neq 0}$ in the $n_{V}=1$ case:
\begin{equation}
\gamma -1=\mathcal{O}_{non-BPS,Z\neq 0,n_{V}=1}\equiv \frac{g^{-1}\left|
D_{z}Z\right| _{non-BPS,Z\neq 0}^{2}}{\left| Z\right| _{non-BPS,Z\neq 0}^{2}}%
>0.  \label{gamma2}
\end{equation}

\textit{Apriori}, Eqs. (\ref{gamma})-(\ref{gamma2}) do depend on the
particular non-BPS $Z\neq 0$ critical point of $V_{BH}$ being considered,
\textit{i.e.} they are dependent on the particular set of BH charges at
hand, chosen among the BH charge configurations supporting non-BPS $Z\neq 0$
critical points of $V_{BH}$. Put another way, one would \textit{apriori}
conclude that $\mathcal{O}_{non-BPS,Z\neq 0}$ changes its value depending on
which configuration of BH charges is chosen among the ones supposrting
non-BPS $Z\neq 0$ critical points of $V_{BH}$ in $\mathcal{M}_{n_{V}}$ ($%
dim_{\mathbb{C}}\mathcal{M}_{n_{V}}=n_{V}$).

This is not the case for homogeneous symmetric and non-symmetric SKGs, as
respectively computed in \cite{BFGM1} and \cite{DFT07-1}. For such SKGs $%
\gamma =4$ regardless of the peculiar non-BPS $Z\neq 0$ critical point of $%
V_{BH}$ being considered. As claimed in \cite{TT}, $\gamma =4$ seemingly
holds true for every non-BPS $Z\neq 0$ critical point of $V_{BH}$ in generic
$n_{V}$-dim. cubic (not necessarily symmetric, nor homogeneous) SKG.

The strict positivity of $\mathcal{O}_{non-BPS,Z\neq 0}$ (and the subsequent
increasing nature of the multiplicative ``renormalization'' of $%
S_{BH,non-BPS,Z\neq 0}$ with respect to the formal expression of $S_{BH,%
\frac{1}{2}-BPS}$, when it actually occurs) yields that (at least \textit{%
formally}, and in the considered framework) the $\frac{1}{2}$-BPS and
non-BPS $Z\neq 0$ species of critical points of $V_{BH}$ are \textit{%
``discretely disjoint''} one from the other.\setcounter{equation}0

\section{\label{Sect4}Stability of non-BPS Critical Points of $V_{BH}$}

In order to decide whether a critical point of $V_{BH}$ is an attractor in
strict sense, one has to consider the following condition:
\begin{equation}
H_{\mathbb{R}}^{V_{BH}}\equiv H_{ab}^{V_{BH}}\equiv D_{a}D_{b}V_{BH}>0~~~%
\text{at}~~~D_{c}V_{BH}=\frac{\partial V_{BH}}{\partial \phi ^{c}}=0\text{~~~%
}\forall c=1,...,2n_{V},  \label{stab}
\end{equation}
\textit{i.e.} the condition of (strict) positive-definiteness of the real $%
2n_{V}\times 2n_{V}$ Hessian matrix $H_{\mathbb{R}}^{V_{BH}}\equiv
H_{ab}^{V_{BH}}$ of $V_{BH}$ (which is nothing but the squared mass matrix
of the moduli) at the critical points of $V_{BH}$, expressed in the real
parameterization through the $\phi $-coordinates. Since $V_{BH}$ is
positive-definite, a stable critical point (namely, an attractor in strict
sense) is necessarily a(n at least local) minimum, and therefore it fulfills
the condition (\ref{stab}).

In general, $H_{\mathbb{R}}^{V_{BH}}$ may be block-decomposed in $%
n_{V}\times n_{V}$ real matrices:
\begin{equation}
H_{\mathbb{R}}^{V_{BH}}=\left(
\begin{array}{ccc}
\mathcal{A} &  & \mathcal{C} \\
&  &  \\
\mathcal{C}^{T} &  & \mathcal{B}
\end{array}
\right) ,  \label{Hessian-real}
\end{equation}
with $\mathcal{A}$ and $\mathcal{B}$ being $n_{V}\times n_{V}$ real
symmetric matrices:
\begin{equation}
\mathcal{A}^{T}=\mathcal{A},~\mathcal{B}^{T}=\mathcal{B}\Leftrightarrow
\left( H_{\mathbb{R}}^{V_{BH}}\right) ^{T}=H_{\mathbb{R}}^{V_{BH}}.
\end{equation}

In the local complex $\left( z,\overline{z}\right) $-parameterization, the $%
2n_{V}\times 2n_{V}$ Hessian matrix of $V_{BH}$ reads
\begin{equation}
H_{\mathbb{C}}^{V_{BH}}\equiv H_{\widehat{i}\widehat{j}}^{V_{BH}}\equiv
\left(
\begin{array}{ccc}
D_{i}D_{j}V_{BH} &  & D_{i}\overline{D}_{\overline{j}}V_{BH} \\
&  &  \\
D_{j}\overline{D}_{\overline{i}}V_{BH} &  & \overline{D}_{\overline{i}}%
\overline{D}_{\overline{j}}V_{BH}
\end{array}
\right) =\left(
\begin{array}{ccc}
\mathcal{M}_{ij} &  & \mathcal{N}_{i\overline{j}} \\
&  &  \\
\overline{\mathcal{N}_{i\overline{j}}} &  & \overline{\mathcal{M}_{ij}}
\end{array}
\right) ,  \label{Hessian-complex}
\end{equation}
where the hatted indices $\hat{\imath}$ and $\hat{\jmath}$ may be
holomorphic or antiholomorphic. $H_{\mathbb{C}}^{V_{BH}}$ is the matrix
actually computable in the SKG formalism presented in Sect. \ref{SKG-gen}
(see below, Eqs. (\ref{M}) and (\ref{N})). Let us here recall that the
invertible unitary Cayley transformation (\ref{unit-transf}) expresses the
change between the Riemannian $2n_{V}$-dim. $\phi $-parameterization of $%
\mathcal{M}_{n_{V}}$ and the K\"{a}hler $n_{V}$-dim.
holomorphic/antiholomorphic $\left( z,\overline{z}\right) $-parameterization
of $\mathcal{M}_{n_{V}}$, used in previous Sects..

As pointed out above, for SKGs having a globally strict positive-definite
metric tensor $g_{i\overline{j}}$ the condition (\ref{stab}) is
automatically satisfied at the $\frac{1}{2}$-BPS critical points of $V_{BH}$
(defined by Eq. (\ref{BPS-conds})). On the other hand, non-BPS $Z\neq 0$
critical points of $V_{BH}$ does not automatically fulfill the condition (%
\ref{stab}), and a more detailed analysis \cite{BFGM1,AoB-book} is needed.

Using the properties of SKG, one obtains:
\begin{eqnarray}
&&
\begin{array}{l}
\mathcal{M}_{ij}\equiv D_{i}D_{j}V_{BH}=D_{j}D_{i}V_{BH}= \\
\\
=4i\overline{Z}C_{ijk}g^{k\overline{k}}\left( \overline{D}_{\overline{k}}%
\overline{Z}\right) +i\left( D_{j}C_{ikl}\right) g^{k\overline{k}}g^{l%
\overline{l}}\left( \overline{D}_{\overline{k}}\overline{Z}\right) \left(
\overline{D}_{\overline{l}}\overline{Z}\right) ;
\end{array}
\label{M} \\
&&  \notag \\
&&  \notag \\
&&
\begin{array}{l}
\mathcal{N}_{i\overline{j}}\equiv D_{i}\overline{D}_{\overline{j}}V_{BH}=%
\overline{D}_{\overline{j}}D_{i}V_{BH}= \\
\\
=2\left[ g_{i\overline{j}}\left| Z\right| ^{2}+\left( D_{i}Z\right) \left(
\overline{D}_{\overline{j}}\overline{Z}\right) +g^{l\overline{n}}C_{ikl}%
\overline{C}_{\overline{j}\overline{m}\overline{n}}g^{k\overline{k}}g^{m%
\overline{m}}\left( \overline{D}_{\overline{k}}\overline{Z}\right) \left(
D_{m}Z\right) \right] .
\end{array}
\label{N}
\end{eqnarray}
Here we limit ourselves to point out that further noteworthy elaborations of
$\mathcal{M}_{ij}$ and $\mathcal{N}_{i\overline{j}}$ can be performed in
homogeneous symmetric SK manifolds, where $D_{j}C_{ikl}=0$ globally \cite
{BFGM1}, and that the K\"{a}hler-invariant $\left( 2,2\right) $-tensor $g^{l%
\overline{n}}C_{ikl}\overline{C}_{\overline{j}\overline{m}\overline{n}}$ can
be rewritten in terms of the Riemann-Christoffel tensor $R_{i\overline{j}k%
\overline{m}}$ by using the so-called ``SKG constraints'' (see the third of
Eqs. (\ref{C})) \cite{AoB-book}. Moreover, the differential Bianchi
identities for $R_{i\overline{j}k\overline{m}}$ imply $\mathcal{M}_{ij}$ to
be symmetric (see comment below Eqs. (\ref{C})).

Thus, one gets the following global properties:
\begin{equation}
\mathcal{M}^{T}=\mathcal{M},~~\mathcal{N}^{\dag }=\mathcal{N}\Leftrightarrow
\left( H_{\mathbb{C}}^{V_{BH}}\right) ^{T}=H_{\mathbb{C}}^{V_{BH}},
\end{equation}
implying that
\begin{equation}
\left( H_{\mathbb{C}}^{V_{BH}}\right) ^{\dag }=H_{\mathbb{C}%
}^{V_{BH}}\Leftrightarrow \mathcal{M}^{\dag }=\mathcal{M},~~\mathcal{N}^{T}=%
\mathcal{N}\Leftrightarrow \overline{\mathcal{M}}=\mathcal{M},~~\overline{%
\mathcal{N}}=\mathcal{N}.
\end{equation}
It should be stressed clearly that the symmetry but non-Hermiticity of $H_{%
\mathbb{C}}^{V_{BH}}$ actually does not matter, because what one is
ineterested in are the eigenvalues of the real form $H_{\mathbb{R}}^{V_{BH}}$%
, which is real and symmetric, and therefore admitting $2n_{V}$ \textit{real}
eigenvalues.

In order to relate $H_{\mathbb{R}}^{V_{BH}}$ expressed by Eq. (\ref
{Hessian-real}) with $H_{\mathbb{C}}^{V_{BH}}$ given by Eq. (\ref
{Hessian-complex}), we exploit the invertible unitary Cayley transformation (%
\ref{unit-transf}), obtaining the following relations between the $%
n_{V}\times n_{V}$ sub-blocks of $H_{\mathbb{R}}^{V_{BH}}$ and $H_{\mathbb{C}%
}^{V_{BH}}$:
\begin{equation}
\left\{
\begin{array}{l}
\mathcal{M}=\frac{1}{2}\left( \mathcal{A}-\mathcal{B}\right) +\frac{i}{2}%
\left( \mathcal{C}+\mathcal{C}^{T}\right) ; \\
\\
\mathcal{N}=\frac{1}{2}\left( \mathcal{A}+\mathcal{B}\right) +\frac{i}{2}%
\left( \mathcal{C}^{T}-\mathcal{C}\right) ,
\end{array}
\right.  \label{4jan1}
\end{equation}
or its inverse
\begin{equation}
\left\{
\begin{array}{l}
\mathcal{A}=Re\mathcal{M}+Re\mathcal{N}; \\
\\
\mathcal{B}=Re\mathcal{N}-Re\mathcal{M}; \\
\\
\mathcal{C}=Im\mathcal{M}-Im\mathcal{N}.
\end{array}
\right.  \label{4jan2}
\end{equation}
The matrix action of the invertible Cayley unitary transformation (\ref
{unit-transf}) may be encoded in a matrix $\mathcal{U}\in U(2n_{V})$ ($%
\Leftrightarrow \mathcal{U}^{-1}=\mathcal{U}^{\dag }$):
\begin{equation}
H_{\mathbb{R}}^{V_{BH}}=\mathcal{U}^{-1}H_{\mathbb{C}}^{V_{BH}}\left(
\mathcal{U}^{T}\right) ^{-1},  \label{transf-1}
\end{equation}
or equivalently
\begin{equation}
H_{\mathbb{C}}^{V_{BH}}=\mathcal{U}H_{\mathbb{R}}^{V_{BH}}\mathcal{U}^{T}.
\label{transf-2}
\end{equation}

The structure of the Hessian matrix gets simplified at the critical points
of $V_{BH}$, because the covariant derivatives may be substituted by the
flat ones; the critical Hessian matrices in complex
holomorphic/antiholomorphic and real local parameterizations respectively
read
\begin{equation}
\left. H_{\mathbb{C}}^{V_{BH}}\right| _{\partial V_{BH}=0}\equiv \left(
\begin{array}{ccc}
\partial _{i}\partial _{j}V_{BH} &  & \partial _{i}\overline{\partial }_{%
\overline{j}}V_{BH} \\
&  &  \\
\partial _{j}\overline{\partial }_{\overline{i}}V_{BH} &  & \overline{%
\partial }_{\overline{i}}\overline{\partial }_{\overline{j}}V_{BH}
\end{array}
\right) _{\partial V_{BH}=0}=\left(
\begin{array}{ccc}
\mathcal{M} &  & \mathcal{N} \\
&  &  \\
\overline{\mathcal{N}} &  & \overline{\mathcal{M}}
\end{array}
\right) _{\partial V_{BH}=0};  \label{Hessian-complex-crit}
\end{equation}
\begin{equation}
\left. H_{\mathbb{R}}^{V_{BH}}\right| _{\partial V_{BH}=0}=\left. \frac{%
\partial ^{2}V_{BH}}{\partial \phi ^{a}\partial \phi ^{b}}\right| _{\partial
V_{BH}=0}=\left(
\begin{array}{ccc}
\mathcal{A} &  & \mathcal{C} \\
&  &  \\
\mathcal{C}^{T} &  & \mathcal{B}
\end{array}
\right) _{\partial V_{BH}=0}.  \label{Hessian-real-crit}
\end{equation}
Thus, the study of the condition (\ref{stab}) finally amounts to the study
of the \textit{eigenvalue problem} of the real symmetric $2n_{V}\times
2n_{V} $ critical Hessian matrix $\left. H_{\mathbb{R}}^{V_{BH}}\right|
_{\partial V_{BH}=0}$ given by Eq. (\ref{Hessian-real-crit}), which is the
Cayley-transformed (through Eq. (\ref{transf-1})) of the complex (symmetric,
but not necessarily Hermitian) $2n_{V}\times 2n_{V}$ critical Hessian $%
\left. H_{\mathbb{C}}^{V_{BH}}\right| _{\partial V_{BH}=0}$ given by Eq. (%
\ref{Hessian-complex-crit}).\medskip\

Once again, the situation strongly simplifies in $n_{V}=1$ SKG.

Indeed, for $n_{V}=1$ the moduli-dependent matrices $\mathcal{A}$, $\mathcal{%
B}$, $\mathcal{C}$, $\mathcal{M}$ and $\mathcal{N}$ introduced above are
simply scalar functions. In particular, $\mathcal{N}$ is real, since $%
\mathcal{C}$ trivially satisfies $\mathcal{C}=\mathcal{C}^{T}$. The
stability condition (\ref{stab}) can thus be written as
\begin{equation}
H_{\mathbb{R}}^{V_{BH}}\equiv D_{a}D_{b}V_{BH}>0\text{,~}\left(
a,b=1,2\right) ~~~\text{at}~~~D_{c}V_{BH}=\frac{\partial V_{BH}}{\partial
\phi ^{c}}=0\text{~~~}\forall c=1,2.
\end{equation}
It may be easily shown that such a stability condition for critical points
of $V_{BH}$ in $n_{V}=1$ SKG can be equivalently reformulated as the strict
bound
\begin{equation}
\left. \mathcal{N}\right| _{\partial V_{BH}=0}>\left| \mathcal{M}\right|
_{\partial V_{BH}=0},  \label{stab-1}
\end{equation}
where
\begin{eqnarray}
\mathcal{N} &\equiv &D_{z}\overline{D}_{\overline{z}}V_{BH}=\overline{D}_{%
\overline{z}}D_{z}V_{BH}=2\left[ g\left| Z\right| ^{2}+\left| D_{z}Z\right|
^{2}+\left| C\right| ^{2}g^{-3}\left| D_{z}Z\right| ^{2}\right] ; \\
&&  \notag \\
\mathcal{M} &\equiv &D_{z}D_{z}V_{BH}=4i\overline{Z}Cg^{-1}\left( \overline{D%
}_{\overline{z}}\overline{Z}\right) +i\left( D_{z}C\right) g^{-2}\left(
\overline{D}_{\overline{z}}\overline{Z}\right) ^{2}.
\end{eqnarray}

As it has to be from the treatment given in Sect. \ref{Sect2}, $\frac{1}{2}$%
-BPS critical points of $V_{BH}$ (determined in the $n_{V}=1$ case by the
unique differential condition $D_{z}Z=0$) automatically satisfies the strict
bound (\ref{stab-1}).\smallskip

Let us now consider the non-BPS, $Z\neq 0$ critical points of $V_{BH}$
introduced in Sect. \ref{Sect3}. By evaluating the functions $\mathcal{N}$
and $\mathcal{M}$ at such a class of points and using the second of
relations (\ref{rel1}), one gets\cite{BFM}
\begin{equation}
\left. \mathcal{N}\right| _{non-BPS,Z\neq 0}=2\left[ \left| D_{z}Z\right|
^{2}\left( 1+\frac{5}{4}\left| C\right| ^{2}g^{-3}\right) \right]
_{non-BPS,Z\neq 0};  \label{N-non-BPS-Z<>0}
\end{equation}
\begin{equation}
\left| \mathcal{M}\right| _{non-BPS,Z\neq 0}^{2}=4\left\{ \left|
D_{z}Z\right| ^{4}\left[ \left| C\right| ^{4}g^{-6}+\frac{1}{4}g^{-4}\left|
D_{z}C\right| ^{2}+2g^{-3}Re\left[ C\left( \overline{D}_{\overline{z}}%
\overline{C}\right) \left( \overline{D}_{\overline{z}}ln\overline{Z}\right) %
\right] \right] \right\} _{non-BPS,Z\neq 0}.  \label{M-non-BPS-Z<>0}
\end{equation}
By substituting Eqs. (\ref{N-non-BPS-Z<>0}) and (\ref{M-non-BPS-Z<>0}) into
the strict inequality Eq. (\ref{stab-1}), one finally obtains the stability
condition for non-BPS, $Z\neq 0$ critical points of $V_{BH}$ in $n_{V}=1$
SKG:
\begin{gather}
\left. \mathcal{N}\right| _{non-BPS,Z\neq 0}>\left| \mathcal{M}\right|
_{non-BPS,Z\neq 0}; \\
\Updownarrow  \notag \\
1+\frac{5}{4}\left( \left| C\right| ^{2}g^{-3}\right) _{non-BPS,Z\neq 0}>
\notag \\
>\sqrt{\left[ \left| C\right| ^{4}g^{-6}+\frac{1}{4}g^{-4}\left|
D_{z}C\right| ^{2}+2g^{-3}Re\left[ C\left( \overline{D}_{\overline{z}}%
\overline{C}\right) \left( \overline{D}_{\overline{z}}ln\overline{Z}\right) %
\right] \right] _{non-BPS,Z\neq 0}}.  \notag  \label{stab-1-1}
\end{gather}

It is immediate to notice that Eq. (\ref{stab-1-1}) is satisfied for sure
when the function $C$ is globally covariantly constant, \textit{i.e.} when $%
D_{z}C=0$ globally \cite{CKV,CVP}. Because of the fact that $\forall
n_{V}\in \mathbb{N}$ quadratic (homogeneous symmetric) SKGs does not admit
non-BPS, $Z\neq 0$ critical points of $V_{BH}$ \cite{BFGM1}, the $n_{V}=1$
homogeneous symmetric SKG automatically satisfying the condition (\ref
{stab-1-1}) corresponds to the SK manifold\textbf{\ }$\frac{SU(1,1)}{U(1)}$,
endowed with a cubic holomorphic prepotential which (in a suitable system of
local special symplectic coordinates) reads
\begin{equation}
F(t)=\lambda t^{3},~\lambda \in \mathbb{C},
\end{equation}
and constrained by the condition $Im\left( t\right) <0$ (usually in the
literature $\lambda =\frac{1}{3}$, but such a choice does not yield any loss
of generality).

Such an $n_{V}=1$ SKG may be obtained by putting $n=-2$ in the so-called
cubic reducible rank-$3$ infinite sequence of homogeneous symmetric SK
manifolds $\frac{SU(1,1)}{U(1)}\otimes \frac{SO(2,2+n)}{SO(2)\otimes SO(2+n)}
$ (where $n_{V}=n+rank=n+3$). It should be noticed that $\frac{SU(1,1)}{U(1)}
$ endowed with $F(t)=\lambda t^{3}$ actually is the ``smallest'' element of
the infinite family $\frac{SU(1,1)}{U(1)}\otimes \frac{SO(2,2+n)}{%
SO(2)\otimes SO(2+n)}$, which indeed does \textit{not} admit the $n_{V}=0$
case, \textit{i.e.} the pure $\mathcal{N}=2$, $d=4$ supergravity theory, as
a limit case\footnote{%
The only homogeneous symmetric SKG admitting a consistent (and obtained by
vanishing some moduli) $n_{V}=0$ limit (reached for $n=-1$) is the quadratic
one of the irreducible rank-$1$ infinite sequence $\frac{SU(1,1+n)}{%
U(1)\otimes SU(1+n)}$ (see \cite{BFGM1} and Refs. therein). The homogeneous
non-symmetric SKGs (see \textit{e.g.} \cite{DFT07-1} and Refs. therein),
because of they all are cubic, do not admit a consistent (and obtained by
vanishing some moduli) $n_{V}=0$ limit.} (pure supergravity would indeed be
reached by putting $n=3$, but such a case is not admitted).

Moreover, the manifold $\frac{SU(1,1)}{U(1)}$ endowed with $F(t)=\lambda
t^{3}$ corresponds to nothing but a peculiar \textit{%
triality-symmetry-destroying degeneration} of the noteworthy $n_{V}=3$ $stu$
SKG, based on the manifold $\left( \frac{SU(1,1)}{U(1)}\right) ^{3}$ and
endowed with a cubic holomorphic prepotential which (in a suitable system of
manifestly triality-invariant\footnote{%
The noteworthy \textit{triality symmetry} of the $stu$ $n_{V}=3$ SKG
has been recently related to quantum information theory \cite
{Duff}-\nocite{KL,Levay,Ferrara-Duff1,Levay2}\cite {Ferrara-Duff2}.}
local special symplectic coordinates) reads $F(s,t,u)=stu$
\cite{BKRSW} (see also \cite {Shmakova} and \cite{K3}). Indeed,
$F(t)=\lambda t^{3}$ can be obtained from $F(s,t,u)=stu$
\textit{e.g.} by identifying $s=t=u$, and by a further
suitable rescaling, \textit{e.g.} by rescaling every modulus by $\lambda ^{-%
\frac{1}{3}}$\ (in the choice $\lambda =\frac{1}{3}$, rescaling by $\sqrt[3]{%
3}$).

It should be also pointed out that the $n_{V}=1$-dim. (in $\mathbb{C}$) SK
manifold $\frac{SU(1,1)}{U(1)}$ can also be obtained as the $n=0$ element of
the quadratic irreducible rank-$1$ infinite sequence $\frac{SU(1,1+n)}{%
U(1)\otimes SU(1+n)}$, but in such a case it would be endowed with a
quadratic holomorphic prepotential function reading - in a suitable system
of local special symplectic coordinates - $F(t)=\frac{i}{4}\left(
t^{2}-1\right) $ (see \cite{BFGM1} and Refs. therein). Such differences at
the level of prepotential determine actual different geometrical properties.
For instance, by working in a suitable system of local special symplectic
coordinates and using the first and third of Eqs. (\ref{C}), one obtains the
following values for the scalar curvature $R$:
\begin{equation}
\begin{array}{l}
\frac{SU(1,1)}{U(1)},F(t)=\lambda t^{3}:R\equiv g^{-2}R_{1\overline{1}1%
\overline{1}}=-\frac{2}{3}; \\
\\
\frac{SU(1,1)}{U(1)},F(t)=\frac{i}{4}\left( t^{2}-1\right) :R\equiv
g^{-2}R_{1\overline{1}1\overline{1}}=-2,
\end{array}
\end{equation}
where $R_{1\overline{1}1\overline{1}}$ denotes the unique component of the
Riemann tensor in $n_{V}=1$ (S)KG, and the global values $C=0$ for the
quadratic case and\footnote{%
The global value $\left| C\right| ^{2}g^{-3}=\frac{4}{3}$ for homogeneous
symmetric cubic $n_{V}=1$ SKGs actually is nothing but the $n_{V}=1$ case of
the general global relation holding in a generic homogeneous symmetric cubic
$n_{V}$-dimensional SKG \cite{CVP,GST2}:
\begin{equation*}
C_{p(kl}C_{ij)n}g^{n\overline{n}}g^{p\overline{p}}\overline{C}_{\overline{n}%
\overline{p}\overline{m}}=C_{\left( p\right| (kl}C_{ij)\left| n\right) }g^{n%
\overline{n}}g^{p\overline{p}}\overline{C}_{\overline{n}\overline{p}%
\overline{m}}=\frac{4}{3}g_{\left( l\right| \overline{m}}C_{\left|
ijk\right) }.
\end{equation*}
} $\left| C\right| ^{2}g^{-3}=\frac{4}{3}$ for the cubic case were
respectively used.

Clearly, the cubic homogeneous symmetric $n_{V}=1$ SKG based on $\frac{%
SU(1,1)}{U(1)}$ is not the only one admitting non-BPS, $Z\neq 0$ critical
points satisfying the stability condition (\ref{stab-1-1}). In the general
case $\left( D_{z}C\right) _{non-BPS,Z\neq 0}$ is the fundamental
geometrical quantity playing a key role in determining the stability of
non-BPS, $Z\neq 0$ critical points of $V_{BH}$ in $n_{V}=1$ SKG.

\setcounter{equation}0

\section{\label{Conclusion}\textbf{Further Results, Some Developments and
Outlook}}

The present report dealt with some recent advances in the study of extremal
BH attractors in $\mathcal{N}=2$, $d=4$ MESGT.

We discussed the AEs for a generic number $n_{V}$ of moduli in a static,
spherically symmetric, asymptotically flat extremal BH background. Such Eqs.
are nothing but the criticality conditions for a real, positive-definite
``effective BH potential'' function $V_{BH}$ defined on the SK vector
supermultiplets' scalar manifold $\mathcal{M}_{n_{V}}$.

$V_{BH}$ is one of the two invariants of the SK geometry of $\mathcal{M}%
_{n_{V}}$ which are quadratic (and thus lowest-order) in the BH charges,
defined as the electric and magnetic fluxes of the field-strength two-forms
of the $n_{V}+1$ Maxwell vector fields of the $\mathcal{N}=2$, $d=4$ MESGT
being considered ($n_{V}$ is the number of Abelian vector multiplets, and
also the graviphoton from the supergravity multiplet has to be taken into
account).

Due to staticity and spherical symmetry, the (bosonic sector of the)
considered $\mathcal{N}=2$, $d=4$ MESGT can be described by an effective $1$%
-dimensional Lagrangian in the radial (evolution) variable. Peculiar
features of a spontaneously broken supergravity theory arise in such a
Lagrangian effective formalism, in which the condition of existence of
non-BPS critical points of $V_{BH}$ (with non-vanishing central charge $Z$)
is given by the vanishing of the determinant of a (fermionic) gaugino mass
matrix.

Concerning the stability of the critical points of $V_{BH}$, because of $%
V_{BH}$ is positive-definite, they necessarily must be (at least local)
minima in order to correspond to \textit{attractor} horizon scalar
configurations in strict sense. In general, the stability is controlled by
the SKG of $\mathcal{M}_{n_{V}}$: in addition to the rank-$3$, completely
symmetric, covariantly holomorphic tensor $C_{ijk}$, also its covariant
derivatives $D_{i}C_{jkl}$ (related, through the so-called \textit{SK
geometry contraints}, to the covariant derivatives $D_{m}R_{i\overline{j}k%
\overline{l}}$ of the Riemann-Christoffel tensor) turn out to be crucial.
This can easily be seen by considering the explicit expression of $H_{%
\mathbb{C}}^{V_{BH}}$, the $2n_{V}\times 2n_{V}$ covariant Hessian matrix in
the complex holomorphic/antiholomorphic parameterization of $\mathcal{M}%
_{n_{V}}$. In order to decide whether a critical point of $V_{BH}$ actually
gives rise to an attractor in strict sense, one has actually to study the
eigenvalue problem for $H_{\mathbb{R}}^{V_{BH}}$, real form of $H_{\mathbb{C}%
}^{V_{BH}}$, properly evaluated at the considered critical point.

The so-called $\frac{1}{2}$-BPS critical points of $V_{BH}$ (treated in
Sect. \ref{Sect2}) correspond to horizon scalar configurations which
preserve half of the supersymmetry degrees of freedom of the asymptotical
Minkowski background (namely, 4 out of 8). They are always stable, thus
corresponding to \textit{attractors} in strict sense. Other two species of
critical points of $V_{BH}$ exist in $\mathcal{N}=2$, $d=4$ MESGT, \textit{%
i.e.} the non-BPS $Z\neq 0$ (treated in Sects. \ref{Sect3} and \ref{Sect4})
and non-BPS $Z=0$ ones. In general, both such classes of critical points are
not necessarily stable; the condition(s) for their stability can be
formulated in purely geometrical terms, by using the properties of the SKG
of $\mathcal{M}_{n_{V}}$.

As it happens for the study of SKG, also the eigenvalue problem of $H_{%
\mathbb{R}}^{V_{BH}}$ strongly simplifies in the case $n_{V}=1$, \textit{i.e.%
} in the case in which only $1$ Maxwell vector multiplet is coupled to the
supergravity multiplet. Consequently, only $2$ Abelian vector fields are
present in such a case: the graviphoton one and the one coming from the
unique Abelian supermultiplet. The stability condition for non-BPS, $Z\neq 0$
critical points of $V_{BH}$ in a generic $n_{V}=1$ SKG can be shown to be
equivalent to a strict inequality, involving the fundamental geometrical
entities of the SKG of $\mathcal{M}_{n_{V}=1}$ (this actually happens also
for the non-BPS, $Z=0$ case \cite{BFMY}, not treated in the present report).

Recently, in \cite{BFGM1} the general solutions to the AEs were obtained and
classified by group-theoretical methods for the peculiar class of $\mathcal{N%
}=2$, $d=4$ MESGTs having an \textit{homogeneous symmetric} SK scalar
manifold, \textit{i.e.} for those $\mathcal{N}=2$, $d=4$ MESGTs in which $%
\mathcal{M}_{n_{V}}$, beside being SK, is a coset $\frac{G}{H}$ with a
globally covariantly constant Riemann-Christoffel tensor $R_{i\overline{j}k%
\overline{l}}$: $D_{m}R_{i\overline{j}k\overline{l}}=0$. Such a conditions
can be transported on $C_{ijk}$ by means of the so-called \textit{SK
geometry contraints}, obtaining: $D_{l}C_{ijk}=0$.

The considered $\mathcal{N}=2$, $d=4$ MESGTs are usually named \textit{%
homogeneous symmetric} MESGTs, and they have been classified in literature
\cite{CKV,CVP,dWVVP}.

With the exception of the ones based on\footnote{%
The quadratic irreducible rank-$1$ infinite sequence $\frac{SU(1,1+n)}{%
U(1)\otimes SU(1+n)}$ has $C_{ijk}=0$ globally. As shown in App. I of \cite
{BFGM1}, such a family has only two classes of non-degenerate solutions to
the AEs: one $\frac{1}{2}$-BPS and one non-BPS with $Z=0$.} $\frac{SU(1,1+n)%
}{U(1)\otimes SU(1+n)}$, \textit{all} homogeneous symmetric SKGs are endowed
with cubic holomorphic prepotentials. In all rank-$3$ homogeneous symmetric
cubic SK manifolds $\frac{G}{H=H_{0}\otimes U(1)}$ (being the vector
supermultiplets' scalar manifolds of $\mathcal{N}=2$, $d=4$ MESGTs defined
by Jordan algebras of degree $3$), the solutions to AEs have been shown to
exist in three distinct classes, one $\frac{1}{2}$-BPS and the other two
non-BPS, one of which corresponds to vanishing central charge $Z=0$. It is
here worth remarking that the non-BPS $Z=0$ class of solutions to AEs has no
analogue in $d=5$, where a similar classification has been recently given
\cite{FG2}.

Furthermore, the three classes of critical points of $V_{BH}$ in $\mathcal{N}%
=2$, $d=4$ homogeneous symmetric cubic MESGTs have been put in \textit{%
one-to-one} correspondence with the non-degenerate charge orbits of the
actions of the $U$-duality groups $G$ on the corresponding BH charge
configuration spaces. In other words, the three species of solutions to AEs
in $\mathcal{N}=2$, $d=4$ homogeneous symmetric cubic MESGTs are supported
by configurations of the BH charges lying along the non-degenerate
typologies of charge orbits of the $U$-duality group $G$ in the real
(electric-magnetic field strengths) representation space $R_{V}$. The
results obtained in \cite{BFGM1} are summarized in Table 1.

\begin{table}[t]
\begin{center}
\begin{tabular}{|c||c|c|c|}
\hline
& $
\begin{array}{c}
\\
\frac{1}{2}\text{-BPS orbits } \\
~~\mathcal{O}_{\frac{1}{2}-BPS}=\frac{G}{H_{0}} \\
~
\end{array}
$ & $
\begin{array}{c}
\\
\text{non-BPS, }Z\neq 0\text{ orbits} \\
\mathcal{O}_{non-BPS,Z\neq 0}=\frac{G}{\widehat{H}}~ \\
~
\end{array}
$ & $
\begin{array}{c}
\\
\text{non-BPS, }Z=0\text{ orbits} \\
\mathcal{O}_{non-BPS,Z=0}=\frac{G}{\widetilde{H}}~ \\
~
\end{array}
$ \\ \hline\hline
$
\begin{array}{c}
\\
\text{Quadratic} \\
\text{~Sequence}
\end{array}
$ & $\frac{SU(1,n+1)}{SU(n+1)}~$ & $-$ & $\frac{SU(1,n+1)}{SU(1,n)}~$ \\
\hline
$
\begin{array}{c}
\\
\text{Cubic} \\
\text{Sequence~}
\end{array}
$ & $\frac{SU(1,1)\otimes SO(2,2+n)}{SO(2)\otimes SO(2+n)}~$ & $\frac{%
SU(1,1)\otimes SO(2,2+n)}{SO(1,1)\otimes SO(1,1+n)}~$ & $\frac{%
SU(1,1)\otimes SO(2,2+n)}{SO(2)\otimes SO(2,n)}$ \\ \hline
$
\begin{array}{c}
\\
J_{3}^{\mathbb{O}} \\
~
\end{array}
$ & $\frac{E_{7(-25)}}{E_{6}}$ & $\frac{E_{7(-25)}}{E_{6(-26)}}$ & $\frac{%
E_{7(-25)}}{E_{6(-14)}}~$ \\ \hline
$
\begin{array}{c}
\\
J_{3}^{\mathbb{H}} \\
~
\end{array}
$ & $\frac{SO^{\ast }(12)}{SU(6)}~$ & $\frac{SO^{\ast }(12)}{SU^{\ast }(6)}~$
& $\frac{SO^{\ast }(12)}{SU(4,2)}~$ \\ \hline
$
\begin{array}{c}
\\
J_{3}^{\mathbb{C}} \\
~
\end{array}
$ & $\frac{SU(3,3)}{SU(3)\otimes SU(3)}$ & $\frac{SU(3,3)}{SL(3,\mathbb{C})}$
& $\frac{SU(3,3)}{SU(2,1)\otimes SU(1,2)}~$ \\ \hline
$
\begin{array}{c}
\\
J_{3}^{\mathbb{R}} \\
~
\end{array}
$ & $\frac{Sp(6,\mathbb{R})}{SU(3)}$ & $\frac{Sp(6,\mathbb{R})}{SL(3,\mathbb{%
R})}$ & $\frac{Sp(6,\mathbb{R})}{SU(2,1)}$ \\ \hline
\end{tabular}
\end{center}
\caption{\textbf{Non-degenerate orbits of $\mathcal{N}=2$, $d=4$ homogeneous
symmetric MESGTs }}
\end{table}

In all the $\mathcal{N}=2$, $d=4$ homogeneous symmetric MESGTs based on rank-%
$3$ SK cubic manifolds, the classical BH entropy is given by the
Bekenstein-Hawking entropy-area formula \cite{BH1}
\begin{equation}
S_{BH}=\frac{A_{H}}{4}=\pi \left. V_{BH}\right| _{\partial V_{BH}=0}=\pi
\sqrt{\left| I_{4}\right| },
\end{equation}
where $I_{4}$ is the (lowest order, quartic\footnote{%
For the quadratic irreducible rank-$1$ infinite sequence $\frac{SU(1,1+n)}{%
U(1)\otimes SU(1+n)}$ the lowest-order $G$-invariant is instead quadratic in
the BH charges; it is positive for $\frac{1}{2}$-BPS orbits and negative for
the non-BPS ($Z=0$) ones (see App. I of \cite{BFGM1}).} in the BH charges)
moduli-independent $G$-invariant built out of the (considered non-degenerate
charge orbit in the) representation $R_{V}$. $\frac{1}{2}$-BPS and non-BPS $%
Z=0$ classes have $I_{4}>0$, while the non-BPS $Z\neq 0$ class is
characterized by $I_{4}<0$.

The critical mass spectra split in different ways, depending on the
considered class of non-degenerate charge orbits. In general, both at
non-BPS $Z\neq 0$ and at non-BPS $Z=0$, the critical Hessian matrix (\ref
{crit-Hessian-gen}) usually exhibit zero modes (\textit{i.e.} ``flat''
directions), whose actual attractor nature seemingly further depends on
additional conditions on the charge vector $\widetilde{\Gamma }$, other than
the ones given by the extremality conditions (\ref{crit-cond-gen}) (see
\textit{e.g.} \cite{TT}).\smallskip

An interesting direction for further investigations concerns the study of
extremal BH attractors in more general, \textit{non-cubic} SK geometries. A
noteworthy example is given by the SKGs of the scalar manifolds of those $%
\mathcal{N}=2$, $d=4$ MESGTs obtained as effective, low-energy theories of $%
d=10$ Type IIB superstrings compactified on Calabi-Yau threefolds ($CY_{3}$%
s), away from the limit of large volume of $CY_{3}$s.

Recently, \cite{BFMY} studied the extremal BH attractors in $n_{V}=1$\ SKGs
obtained by compactifications (away from the limit of large volume of the
internal manifold) on a peculiar class of $CY_{3}$s, given by the so-called
(mirror) \textit{Fermat} $CY_{3}$s. Such threefolds are classified by the
\textit{Fermat parameter} $k=5,6,8,10$, and they were firstly found in \cite
{Strom-Witten}. The fourth order linear Picard-Fuchs (PF) ordinary
differential Eqs. determining the holomorphic fundamental period $4\times 1$%
\ vector for such a class of $1$-modulus $CY_{3}$s were found some time ago
for $k=5$\ in \cite{CDLOGP1,CDLOGP2} (see also \cite{Cadavid-Ferrara}), and
for $k=6,8,10$\ in \cite{KT}.

More specifically, \cite{BFMY} dealt with\textbf{\ }the so-called \textit{%
Landau-Ginzburg} (LG) extremal BH attractors, \textit{i.e. }the solutions to
the AEs near the origin $z=0$\ (named \textit{LG point}) of the moduli space
$\mathcal{M}_{n_{V}=1}$ ($dim_{\mathbb{C}}\mathcal{M}_{n_{V}=1}=1$), and the
BH charge configurations supporting $z=0$ to be a critical point of $V_{BH}$
were explicitly determined, as well.

An intriguing development in such a framework would amount to extending to
the \textit{Fermat} $CY_{3}$-compactifications (away from the limit of large
volume of the threefold) the conjecture formulated in Sect. 5 of \cite{K3}.
The conjecture was formulated in the framework of (the large volume limit of
$CY_{3}$-compactifications leading to) the previously mentioned \textit{%
triality-symmetric} cubic $stu$ model \cite{BKRSW,Shmakova,K3}, and it
argues that the instability of the considered non-BPS ($Z\neq 0$) critical
points of $V_{BH}$ might be only apparent, since such attractors might
correspond to multi-centre stable attractor solutions (see also \cite
{Ferrara-Gimon} and Refs. therein), whose stable nature should be
``resolved'' only at sufficiently small distances. The extension of such a
tempting conjecture to non-BPS extremal BH LG attractors in \textit{Fermat} $%
CY_{3}$-compactifications would be interesting; in particular, the extension
to the non-BPS $Z=0$ case might lead to predict the existence (at least in
the considered peculiar $n_{V}=1$ framework) of \textit{non-BPS lines of
marginal stability} \cite{Denef1,Denef2} with $Z=0$.\smallskip

Moreover, it should be here recalled that the PF Eqs. of \textit{Fermat} $%
CY_{3}$s (\cite{CDLOGP1}-\nocite{CDLOGP2, Cadavid-Ferrara}\cite{KT},
see also\cite{BFMY})
exhibit other two species of \textit{regular singular} points, namely the $k$%
-th roots of unity ($z^{k}=1$, the so-called \textit{conifold points}) and
the \textit{point at infinity} $z\longrightarrow \infty $ in the moduli
space, corresponding to the so-called \textit{large complex structure
modulus limit}. Thus, it would be interesting to solve the AEs in proximity
of such regular singular points, \textit{i.e.} it would be worth
investigating \textit{extremal BH conifold attractors} and \textit{extremal
BH large complex structure attractors} in the moduli space of 1-modulus (%
\textit{Fermat}) $CY_{3}$s. Such an investigation would be of interest, also
in view of recent studies of extremal BH attractors in peculiar examples of $%
n_{V}=2$-moduli $CY_{3}$-compactifications \cite{Misra1}.\medskip

Despite the considerable number of papers written on the Attractor Mechanism
in the extremal BHs of the supersymmetric theories of gravitation along the
last years, still much remains to be discovered along the way leading to a
deep understanding of the inner dynamics of (eventually extended) space-time
singularities in supergravities, and hopefully in their fundamental
high-energy counterparts, such as $d=10$ superstrings and $d=11$ $M$%
-theory.\smallskip

\section*{\textbf{Acknowledgments}}

The work of S.B. has been supported in part by the European Community Human
Potential Program under contract MRTN-CT-2004-005104 \textit{``Constituents,
fundamental forces and symmetries of the universe''}.

The work of S.F. has been supported in part by European Community Human
Potential Program under contract MRTN-CT-2004-005104 \textit{``Constituents,
fundamental forces and symmetries of the universe''} and the contract
MRTN-CT-2004-503369 \textit{``The quest for unification: Theory Confronts
Experiments''}, in association with INFN Frascati National Laboratories and
by D.O.E. grant DE-FG03-91ER40662, Task C.

The work of A.M. has been supported by a Junior Grant of the \textit{%
``Enrico Fermi''} Center, Rome, in association with INFN Frascati National
Laboratories.

\end{document}